\documentclass[aps,prd,nofootinbib,twocolumn,superscriptaddress,preprintnumbers,balancelastpage,longbibliography]{revtex4-2}
\usepackage{aas_macros}

\usepackage{placeins}
\usepackage{amsmath,amssymb,mathtools,bm}
\usepackage{graphicx, color, hepunits}
\usepackage[dvipsnames]{xcolor}
\usepackage{cancel}
\usepackage[normalem]{ulem}
\usepackage{float}
\usepackage{filecontents}
\usepackage{multirow}
 \usepackage{hyperref} %Automatically links \label and \ref commands; Always load last
\hypersetup{
    colorlinks=true,       % false: boxed links; true: colored links
    linkcolor=blue,        % color of internal links
    citecolor=blue,        % color of links to bibliography
    filecolor=magenta,     % color of file links
    urlcolor=blue          % color of external links
}
\usepackage[utf8]{inputenc}
\usepackage[english]{babel}
\usepackage{array}

\usepackage{tikz}
\usepackage{tikz-feynman}
\tikzfeynmanset{compat=1.1.0}

\newcommand{\gagg}{g_{a \gamma \gamma}}

\newcommand{\gann}{g_{ann}}
\newcommand{\gapp}{g_{app}}

\newcommand{\ganneffgagg}{g_{ann}^{\rm eff} \times g_{a \gamma \gamma} }

\newcommand{\Fe}{^{57}\mathrm{Fe}}
\newcommand{\Ni}{^{61}\mathrm{Ni}}
\newcommand{\Ge}{^{73}\mathrm{Ge}}

\newcommand{\ganngagg}{g_{ann}\times g_{a \gamma \gamma}}

\newcommand{\es}[2] {\begin{equation} \label{#1} \begin{split} #2 \end{split} \end{equation}}

\newcommand{\mycomment}[1]{}

\begin{document}
\title{Axion lines from nuclear de-excitations in galactic stellar populations}

\preprint{CERN-TH-2025-162, N3AS-25-014}

\author{Orion Ning}
\affiliation{Berkeley Center for Theoretical Physics, University of California, Berkeley, CA 94720, U.S.A.}
\affiliation{Theoretical Physics Group, Lawrence Berkeley National Laboratory, Berkeley, CA 94720, U.S.A.}

\author{Anupam Ray}
\affiliation{Department of Physics, University of California, Berkeley, CA 94720, U.S.A.}

\author{Benjamin R. Safdi}
\affiliation{Berkeley Center for Theoretical Physics, University of California, Berkeley, CA 94720, U.S.A.}
\affiliation{Theoretical Physics Group, Lawrence Berkeley National Laboratory, Berkeley, CA 94720, U.S.A.}
\affiliation{Theoretical Physics Department, CERN, 1211 Geneva, Switzerland}

\date{\today}

\begin{abstract}
We show that mono-energetic axions are produced in abundance through nuclear de-excitations in nearby galaxies such as M87, which is the central galaxy of the Virgo cluster, and the starburst galaxy M82. If the axion couples to both nucleons and photons and is ultralight, then monochromatic hard X-ray signatures are induced by the subsequent axion-to-photon conversion in the magnetic fields permeating these systems. We search for evidence of such signals using NuSTAR data, focusing specifically on the $\Fe$ de-excitation line at 14.4 keV, and we catalog other potentially relevant nuclear lines. We find no evidence for axions from M87 or M82 and set leading constraints on the combined axion-nucleon and axion-photon coupling at the level of $|g_{ann} \times g_{a\gamma \gamma}| \lesssim 1.1 \times 10^{-22}$ GeV$^{-1}$ in the limit $m_a \lesssim 10^{-10}$ eV, at 95\% confidence.
\end{abstract}
\maketitle

\section{Introduction}

Axions may solve the strong-CP problem and provide a viable candidate for dark matter (DM)~\cite{Peccei:1977hh,Peccei:1977ur,Weinberg:1977ma,Wilczek:1977pj,Preskill:1982cy,Abbott:1982af,Dine:1982ah}. Further, they are generically predicted in string theory, where they arise as the zero modes of higher-dimensional gauge fields upon compactification~\cite{Witten:1984dg,Choi:1985je,Barr:1985hk,Svrcek:2006yi,Arvanitaki:2009fg}.  Axions may be classified into the quantum chromodynamics (QCD) axion, whose mass arises predominantly from QCD contributions, and axion-like particles, whose masses are unrelated to QCD.  Ultra-light axion masses, well below {\it e.g.} $10^{-10}$ eV, are natural and indeed expected in string theory constructions since in these cases the mass contributions must be non-local and are exponentially suppressed by instanton actions (see, {\it e.g.},~\cite{Demirtas:2018akl,Halverson:2019cmy,Mehta:2021pwf,Gendler:2023kjt,Benabou:2025kgx} for explicit constructions).  While ultra-light axions do not interact with QCD (but see~\cite{Hook:2018jle}), they are expected to have interactions with quantum electrodynamics (QED) and Standard Model (SM) fermions, including quarks (see~\cite{Hook:2018dlk,DiLuzio:2020wdo,Safdi:2022xkm,OHare:2024nmr} for reviews). Below the QCD confinement scale the axion-quark interactions induce derivative couplings of the axions to nucleons.  As we show in this work, this allows axions to be produced in abundance through nuclear de-excitations in hot, massive stars found in nearby galaxies.  In the case of ultra-light axions, the mono-energetic axions may then convert efficiently to monochromatic X-rays in the magnetic fields in these systems.  We search for these X-rays using NuSTAR data as in~\cite{Ning:2024eky,Ning:2025tit}, which searched for axion-induced hard X-ray signatures in these same galaxies but focusing on emission from Primakoff and electron Compton and bremsstrahlung processes. Unlike in those works, however, the axion-induced signals that we search for here are mono-energetic at known energies. The detection of such a line would thus be a smoking-gun signature of axions. 

\begin{figure}[!t]
\centering
\includegraphics[width=0.49\textwidth]{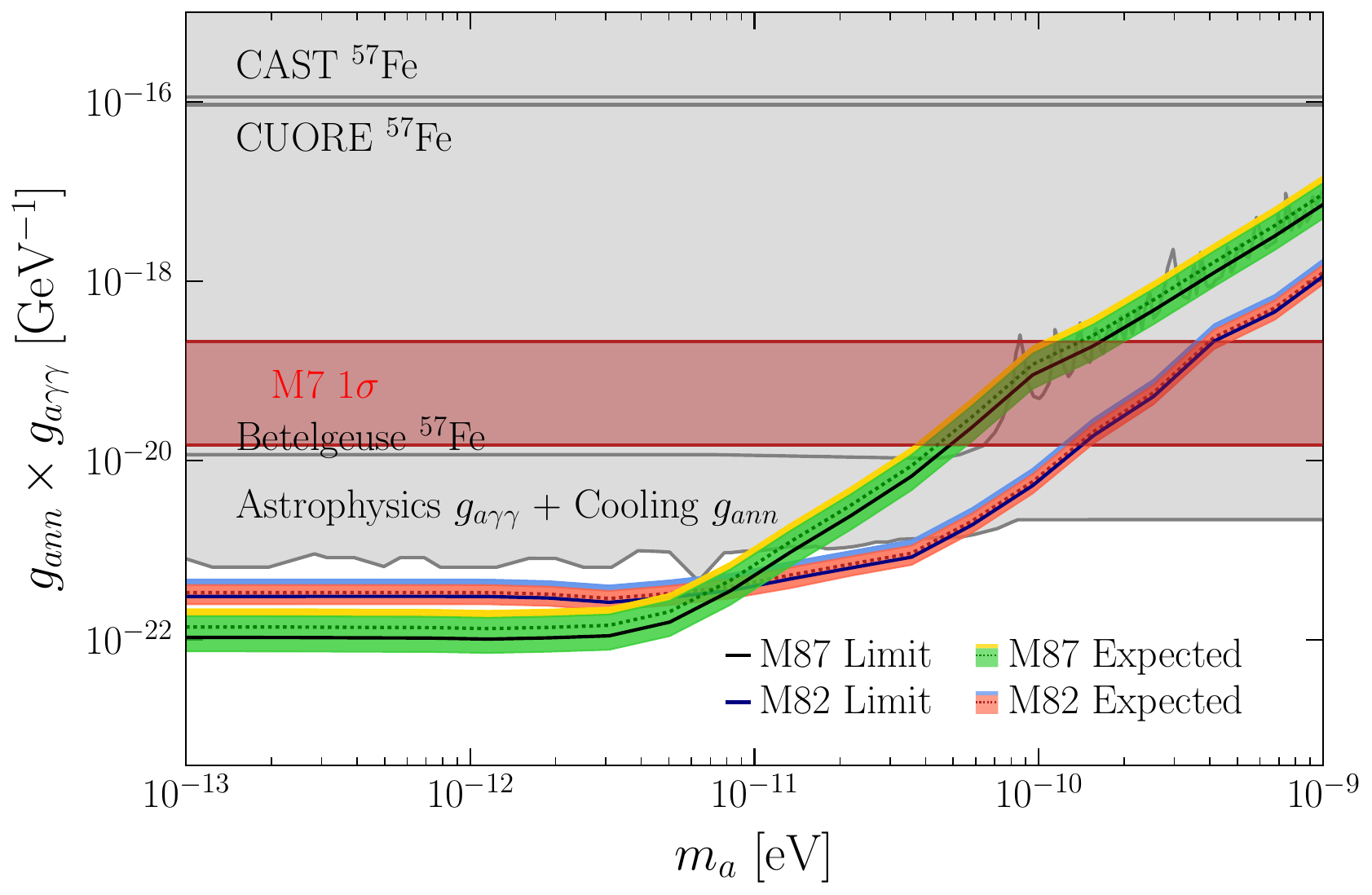}
\vspace{-0.4cm}
\caption{The 95\% upper limits (bolded) on the combined axion-nucleon and axion-photon couplings $\ganngagg$ from our two most stringent searches for ultralight axions produced from nuclear isotope de-excitations in galactic stellar populations. The axions convert to X-rays in the magnetic fields permeating the host galaxies, and we search for these X-rays using NuSTAR data. We also illustrate the $1\sigma/2\sigma$ containment intervals for the expected 95\% upper limits on $\ganngagg$ under the null hypothesis. Our illustrated constraints come from the $\Fe$ searches in M87 and M82, assuming $\gann = \gapp$ (see~\eqref{eq:gann_eff}).  Our results disfavor the axion interpretation of the previously-observed M7 X-ray excess from nearby neutron stars~\cite{Dessert:2019dos,Buschmann:2019pfp} for low mass axions.}
\label{fig:ganngagg}
\end{figure}

It is well established in the context of the Sun that axions may be emitted from the de-excitations of thermally-excited nuclear excited states. The excitation energies of typical nuclei tend to be around $\mathcal{O}$(MeV). Therefore, such emission processes have excitation probabilities that are highly Boltzmann suppressed under normal stellar conditions. However, there exist certain nuclei with low-lying nuclear energy levels in the $\mathcal{O}(10)$ keV range.  Among the nuclei with low-lying nuclear energy levels, $\Fe$, with a transition energy of $E^* = 14.4$ keV, is perhaps the most well-studied~\cite{PhysRevLett.66.2557,Moriyama:1995bz,Krcmar:1998xn,Namba:2007rm,CAST:2009jdc,Derbin:2011zz,CUORE:2012ymr,Armengaud:2013rta,Majorana:2016hop,Avignone:2017ylv,DiLuzio:2021qct,Fleury:2022plh,Candon:2025vpv}.  Yet even this excited state is Boltzmann suppressed at the level of $\sim$$10^{-5}$ in the Sun, which has a core temperature $\sim$1 keV.  On the other hand, we show in this work that accounting for axion production in stellar populations allows us to probe stars with much higher core temperatures, for which the $14.4$ keV $\Fe$ axion line is not Boltzmann suppressed at all.  Our search is similar to that performed in~\cite{Candon:2025vpv}, which searched for the $^{57}$Fe axion line with NuSTAR data towards the nearby supergiant Betelgeuse, though this work achieves superior sensitivity by summing over full stellar populations.

In addition to $\Fe$, previous studies have also explored axion emission and nuclear line searches involving a myriad of other isotopes, each with their own unique line energies, abundances, and other properties, for example, $^{83}\mathrm{Kr}$~\cite{Gavrilyuk:2014mch}, $^{23}\mathrm{Na}$~\cite{PhysRevLett.66.2557,Haxton:2025xqz}, $^{125}\mathrm{Te}$~\cite{Derbin:1997kt}, $^{139}\mathrm{La}$~\cite{Minowa:1993mw}, and $^{65}\mathrm{Cu}$~\cite{PhysRevD.37.618}.  We show in this work that $^{61}$${\rm Ni}$ and $^{73}$${\rm Ge}$ lines at 67.4 keV and $68.8$ keV, respectively, are the next-most luminous emission lines in axions relative to the $\Fe$ line, though the $\Fe$ line dominates in typical stellar populations.

The 14.4 keV $\Fe$ line has been searched for using the CERN Axion Solar Telescope (CAST)~\cite{CAST:2009jdc}, where the solar axions are converted to X-rays in the strong magnetic fields in the CAST experiment (see also~\cite{CUORE:2012ymr} for a related search using the CUORE detector). The upper limit from this search constrains $|\ganneffgagg| \lesssim 10^{-16}$ GeV$^{-1}$, where $g_{a\gamma\gamma}$ is the axion-photon coupling and $g_{ann}^{\rm eff}$ is an effective axion-nucleon coupling (see Fig.~\ref{fig:ganngagg}); these couplings are defined more precisely later in this work.  We improve upon this bound by approximately six orders of magnitude for axion masses less than roughly $10^{-10}$ eV.

Strong constraints also exist independently on the axion-photon and axion-nucleon couplings for low-mass axions; these may be combined to constrain $|\ganngagg|$, as illustrated in Fig.~\ref{fig:ganngagg}. In particular, constraints at the level of $|g_{a\gamma\gamma}| \lesssim 7.8 \times 10^{-13}$ GeV$^{-1}$ for $m_a\lesssim 10^{-10}$ eV at 95\% confidence arise from searching for X-ray signatures with the same NuSTAR data considered in this work towards the nearby active galaxy M82, accounting for axion production in the stellar population in the galaxy through the Primakoff effect and then conversion of axions-to-photons in the magnetic field of the galaxy through inverse Primakoff~\cite{Ning:2024eky}.  Searches for axion-induced spectral modulation in Chandra X-ray data taken towards NGC 1275 are comparable~\cite{Reynolds:2019uqt} (see also~\cite{Marsh:2017yvc,Conlon:2017qcw,Reynes:2021bpe}). 
Above $m_a \gtrsim 10^{-10}$ eV constraints from the non-observation of gamma-rays following SN1987A become prominent~\cite{Brockway:1996yr,Grifols:1996id,Payez:2014xsa,Hoof:2022xbe,Manzari:2024jns}, as are the constraints from axion searches in magnetic white dwarfs (MWD), in particular from axion-induced polarization~\cite{Benabou:2025jcv} (see also~\cite{Dessert:2022yqq, Dessert:2021bkv, Ning:2024ozs}). Two different classes of constraints exist on the axion-nucleon couplings that are nearly degenerate in magnitude. Strong constraints exist at the level of $|g_{ann}| \lesssim 1.3 \times 10^{-9}$ and $|g_{app}| \lesssim 1.5 \times 10^{-9}$ for the axion-neutron and axion-proton couplings, respectively, from neutron star (NS) cooling~\cite{Buschmann:2021juv}.  Similar constraints arise by considering axion production in the proto-NS formed after SN1987A; more strongly-coupled axions would have over-cooled the neutron star and modified the observed neutrino burst~\cite{Raffelt:2006cw,Fischer:2016cyd,Chang:2018rso,Carenza:2019pxu,Carenza:2020cis}.

\section{Nuclear Line Searches}
In stellar environments, a finite axion-nucleon coupling opens a novel channel for axion production through nuclear isotope de-excitations. The key idea is that if the temperature of the star becomes comparable to the nuclear transition energy, the nucleus can become thermally excited, and its subsequent de-excitation can result in axion emission. The emitted axion is monochromatic, with an energy corresponding to the transition energy of the specific nucleus.  Before describing the calculations of these transition rates, we briefly review the axion effective field theory (EFT) in order to set the notation used throughout the rest of this work (see~\cite{Hook:2018dlk,DiLuzio:2020wdo,Safdi:2022xkm,OHare:2024nmr} for in-depth reviews).

At energy scales below the electroweak symmetry breaking scale but above the QCD confinement scale we may parametrize the relevant interactions of the axion EFT by
\es{eq:axion_int}{
{\mathcal L} \supset {1 \over 2 f_a } \sum_q
C_{aqq} \partial^\mu a \bar q \gamma_\mu \gamma^5 q  - {1 \over 4} C_{a\gamma\gamma} {e^2 \over 8 \pi^2 f_a} a F_{\mu \nu} \tilde F^{\mu \nu} \,,
}
where $f_a$ is the axion decay constant that sets the energy scale of the ultraviolet (UV) completion, $a$ is the axion field, and in the first term $C_{aqq}$ is a dimensionless Wilson coefficient and the $q$ stand for the light quarks with masses below the energy scale of the EFT. In the second term $e$ is the electric charge, $C_{a\gamma\gamma}$ is the dimensionless axion-photon coupling, and $F$ is the quantum electrodynamics (QED) field strength. The second term in~\eqref{eq:axion_int} may be written as ${\mathcal L} \supset a g_{a\gamma\gamma} {\bm E} \cdot {\bm B}$, with $g_{a\gamma\gamma} \equiv C_{a\gamma\gamma} e^2 / (8 \pi^2 f_a)$ the dimension-full axion-photon coupling and ${\bm E}$ (${\bm B}$) the electric (magnetic) field.  Below the QCD confinement scale the second term in~\eqref{eq:axion_int} is unchanged but the first must be matched to axion interactions with nucleons and pions, though only the axion-nucleon interactions are relevant here.  In particular, we may write
 \es{}{
 {\mathcal L} \subset {1 \over 2 f_a } \sum_N
 C_{aNN} \partial^\mu a \bar N \gamma_\mu \gamma^5 N \,,
 }
 with $N = n, p$ for neutrons and protons and~\cite{GrillidiCortona:2015jxo}
 \es{}{
 C_{ann} &\approx 0.88 C_{auu} - 0.39 C_{add} \,, \\
 C_{app} &\approx 0.88 C_{add} - 0.39 C_{auu} \,,
 }
 with sub-dominant contributions from $C_{ass}$.
 From the coefficients above we may define the dimensionless axion-nucleon couplings
 \es{}{
 g_{ann} \equiv {C_{ann} m_N \over  f_a} \,, \qquad  g_{app} \equiv {C_{app} m_N \over  f_a} \,,
 }
 with $m_N$ the nucleon mass. 

We now turn to the calculation of the axion emissivity for magnetic dipole (M1) transitions in nuclear isotopes. The axion emission rate per unit mass may be parameterized as~\cite{PhysRevLett.66.2557,CAST:2009jdc}
\begin{equation}
     \dot{\mathcal{N}_a} = \mathcal{N} \, \omega_* \, {\frac{1}{\tau_{m}}} \, \frac{1}{1+\alpha} \frac{\Gamma_a}{\Gamma_{\gamma}}
\label{eq:nuclear_one}
\end{equation}
where $\mathcal{N}$ denotes the number of the isotope of interest per unit mass, $\omega_*$ denotes the probability that the excited state responsible for the axion emission is excited ({\it i.e.}, the excited state occupation fraction), {$\tau_{m}$ is the total mean lifetime} of the excited state,
$\alpha$ is the internal conversion coefficient, and $\Gamma_a / \Gamma_{\gamma}$ is the ratio of the axion decay rate to the photon decay rate.  This ratio may be parameterized as follows~\cite{PhysRevLett.66.2557}: 
\begin{equation}
    \frac{\Gamma_a}{\Gamma_{\gamma}} = \left(\frac{k_a}{k_\gamma}\right)^3\frac{1}{2 \pi \alpha_{\rm EM}} \frac{1}{1+ \delta^2} \left[ \frac{g_{app} (\frac{\beta+1}{2}) + g_{ann} (\frac{\beta-1}{2})}{(\mu_0-0.5) \beta + \mu_1-\eta}\right]^2 \,.
\label{eq:nuclear_two}
\end{equation}
In the above equation, $\alpha_{\rm EM} = 1/137$ is the fine structure constant and $\delta $ is the electric quadrupole (E2) to M1 mixing ratio.  Additionally,
$\mu_0 = (\mu_p + \mu_n)/\mu_N \approx 0.88$ and $\mu_1 = (\mu_p - \mu_n)/\mu_N \approx 4.706$~\cite{ParticleDataGroup:2016lqr}  are the isoscalar and isovector nucleon magnetic moments, with $\mu_p$ ($\mu_n$) denoting the magnetic moment of the proton (neutron) and $\mu_N$ the nuclear magneton.  The coefficients $\beta$ and $\eta$ are nuclear-structure dependent ratios, and which are calculated from the nuclear shell model formalism~\cite{PhysRevLett.66.2557}. Lastly, $k_a = \sqrt{E^2_{\gamma}-m^2_a}$ and $k_{\gamma} = E_{\gamma}$ denote the momenta of the outgoing axions and photons, respectively, for the two decay channels. For ultralight axions, which are the focus in this paper, $k_a/k_{\gamma}$ becomes unity.

For $\Fe$, with a de-excitation energy of $E^* \approx 14.4$ keV, the various nuclear factors are taken to be $\delta \approx 0.002$, $\beta \approx -1.19$, and $\eta \approx 0.8$~\cite{PhysRevLett.66.2557}. This leads to
\begin{equation}
\frac{\Gamma_a}{\Gamma_{\gamma}} \approx 1.83\,(g^{\rm eff}_{ann})^2\,,
\label{eq:fe57_ganneff}
\end{equation}
where we define
\es{eq:gann_eff}{
g^{\rm eff}_{ann} &\equiv g_{ann}  -\left( {\beta + 1 \over 2} \right) (g_{app} + g_{ann} ) \\
&\approx g_{ann} + 0.095 (g_{app} + g_{ann} ) \,.
}
The {total mean lifetime of the excited state is assumed to be $\tau_{m} \approx 141.8$ ns}, with the internal conversion coefficient $\alpha \approx 8.56$~\cite{nudat3}. The occupation fraction at temperature $T$ for the first excited state is
\begin{equation}
\omega_*  = \frac{(2J_1+1) e^{-E_*/T}}{(2J_0+1)+(2J_1+1) e^{-E_*/ T}}
\end{equation}\\
where $J_0 = 1/2$ is the angular momentum of the ground state and $J_1=3/2$ is the angular momenta of the first excited state~\cite{nudat3}.

The parameters $\beta$ and $\eta$ are uncertain in that they depend on non-unique choices of the nuclear shell model calculation, such as the effective interaction Hamiltonian and the effective diagonalization scheme.  Refs.~\cite{Avignone:2017ylv,DiLuzio:2021qct} use different schemes to those we adopt above, yielding  $\beta \approx -1.3065$, and $\eta \approx 1.2054$. With this choice of parameters, one computes $\Gamma_a/\Gamma_{\gamma} \approx 2.42\,(g^{\rm eff}_{ann})^2$, where the effective axion-nucleon coupling is given by $g^{\rm eff}_{ann} \approx 0.15 g_{app} + 1.15 g_{ann}$~\cite{Avignone:2017ylv,DiLuzio:2021qct}.  We choose not to adopt this parameter set in our analysis and instead  use~\eqref{eq:fe57_ganneff} (\textit{i.e.}, as in Ref.~\cite{PhysRevLett.66.2557}), because it leads to more conservative results. Still, this computational uncertainty should be kept in mind when interpreting our results, and we cannot rule out the possibility that $\Gamma_a/\Gamma_{\gamma}$ is, in reality, even lower than we assume. 
 
While $\Fe$ provides the most sensitive probe of axions in our work, we also conduct a search for other potentially-promising isotopes with low-lying energy transition energies $E^*$ accessible via the NuSTAR hard X-ray telescope (\textit{i.e.} $E^* \lesssim 85$ keV). In particular, we find that $\Ni$ and $\Ge$, which have $E^*$ around 67.4 keV and 68.7 keV, respectively, are competitive with $\Fe$ in terms of axion production in particular stellar environments. We detail these findings later in this work, with details described in App.~\ref{app:nuclear}. 

Given a stellar model, which we discuss shortly, the axion-induced line emission rate at energy $E^*$ in units of, \textit{e.g.}, erg/s, which we call $L_a$, is found by integrating over the stellar interior: 
\begin{equation}
    L_a = 4 \pi E^* \int_0^R dr \,r^2 \,  \rho(r) \dot{\mathcal{N}}_a\big(T(r)\big)   \,,
\label{eq:stellar_calc}
\end{equation}
where we make explicit the radial dependence of the stellar temperature. We note that the axion emission spectral shape is not precisely monochromatic at $E^*$ but is broadened by a number of effects, ranging from the finite lifetime of the excited state to the finite temperature of the stellar plasma. In reality, however, quantum and Doppler broadening are negligible compared to the broadening by the finite energy resolution of the NuSTAR telescope (see also~\cite{Candon:2025vpv}). 

Following the procedure in~\cite{Ning:2024eky, Ning:2025tit}, each star in the stellar population of interest is described by a state from dedicated simulations in the Modules for Experiments in Stellar Astrophysics (MESA) code package~\cite{2011ApJS..192....3P, 2013ApJS..208....4P}, which is designed to evolve stars over time, given inputs such as stellar mass and metallicity. MESA then returns radial profiles for the star at a stellar age, providing {\it e.g.} the star's temperature, density, as well as ion and electron number densities, which allows us to calculate~\eqref{eq:stellar_calc} above. On the other hand, the abundances ${\mathcal N}$ of the rare isotopes considered in our work, such as $\Fe$, $\Ni$, and $\Ge$, are primarily informed by the initial metallicity of our galactic population models. These abundances are taken to be mostly solar\footnote{{We note that for the exceptional case of M82 there is evidence that specific heavy elements such as Fe likely have sub-solar abundances~\cite{2020ApJ...904..152L, Origlia:2004}. To account for this, we take the average in~\cite{2020ApJ...904..152L} and set the fiducial abundances to be $\approx$$0.38$ times solar. See Fig.~\ref{fig:M82_abundances} in App.~\ref{app:figures} for an illustration.}}, as described more below. This simplification is justified, as we estimate that the existing primordial abundance of these isotopes make up the dominant contribution to a star's isotope abundance, given that these heavy, evolved isotopes are not substantially produced in stellar interiors for most of the stars' lives (see~\cite{Fleury:2022plh} which verifies this point, but also~\cite{Candon:2025vpv} for another approach). Note that we still attempt to include the production of these isotopes from stellar evolution, roughly estimated by, $\textit{e.g.}$, taking the $^{56}\rm{Fe}$ abundances outputted by MESA and scaling them to $\Fe$ abundances through solar abundance ratios (and analogously for other isotopes). While we reiterate that this is only a rough estimate, a more precise treatment (\textit{e.g.} similar to~\cite{Candon:2025vpv}) would suggest that the relevant s-process nucleosynthesis reactions would only produce slightly more heavy elements like $\Fe$ towards the latest periods of a star's life, which would overall have a minimal impact on large population studies such as the ones considered in this work.

\section{Stellar Populations}

Following and extending the ultralight axion searches in~\cite{Ning:2024eky, Ning:2025tit}, here we choose four different galactic targets to analyze: M82, M87, M31, and the Milky Way Galactic Center (GC).  For each of these targets we calculate the summed axion luminosity at energy $E^*$ from the de-excitation transitions described in the previous section. M82, an archetypal starburst galaxy host to a myriad of young, hot, massive stars, has been shown to be particularly promising in the context of axion searches~\cite{Ning:2024eky, Ning:2025tit, Candon:2024eah}. On the other hand, M87 and M31 provide complementary targets with generally older but larger stellar populations. M87 has the advantage of having significantly stronger magnetic fields ideal for axion-to-photon conversion by nature of being embedded in the Virgo cluster, while M31 is the closest galaxy to our own, around $\sim$785 kpc away. We additionally adopt a simplified model of the Milky Way GC as a point of comparison to these extragalactic targets, which has the advantage of proximity. 

Our stellar population modeling for M82 and M87 is identical to that presented in~\cite{Ning:2024eky}, while the model for M31 is detailed in~\cite{Ning:2025tit}. Briefly, we detail the salient features of these models here. For M82, following~\cite{Ning:2024eky} we take as our fiducial model a stellar population of $N_{\rm tot} \approx 10^{10}$ stars drawn from a modified Salpeter initial mass function (IMF), a `two-burst' star formation history (SFH) across $\sim$$10^7$ years, and an initial metallicity of $Z = 0.02$. For M87, following~\cite{Ning:2024eky} we adopt a canonical Salpeter IMF, an exponential $\tau$ SFH model, $N_{\rm tot} \approx 10^{12}$ stars, and $Z \approx 0.02$. Our model for M31 is similar to that for M87 with slightly differing $N_{\rm tot}$ and $\tau$ (see~\cite{Ning:2025tit}). We also estimate the stellar population of the GC in this work, assuming a similar population as M31 but with a fraction of the number of stars determined by the estimated mass of the GC central bulge population~\cite{GRAVITY:2024tth}. Systematic uncertainties on quantities such as the IMF, SFH, $N_{\rm tot}$, and metallicity are explored in~\cite{Ning:2024eky, Ning:2025tit}; briefly, they involve varying the IMF cutoff, SFH burst and/or decay parameters, total numbers of stars inferred from mass estimates, and metallicity variations, all in accordance with observational uncertainties (see~\cite{Ning:2024eky, Ning:2025tit}). Stars from these stellar population models are drawn from corresponding evolved MESA simulations in stellar mass and age, which are then used to construct our total, summed axion signal, in combination with the formalism for nuclear-transition-induced axions discussed earlier. 

\begin{figure}[!t]
\centering
\includegraphics[width=0.49\textwidth]{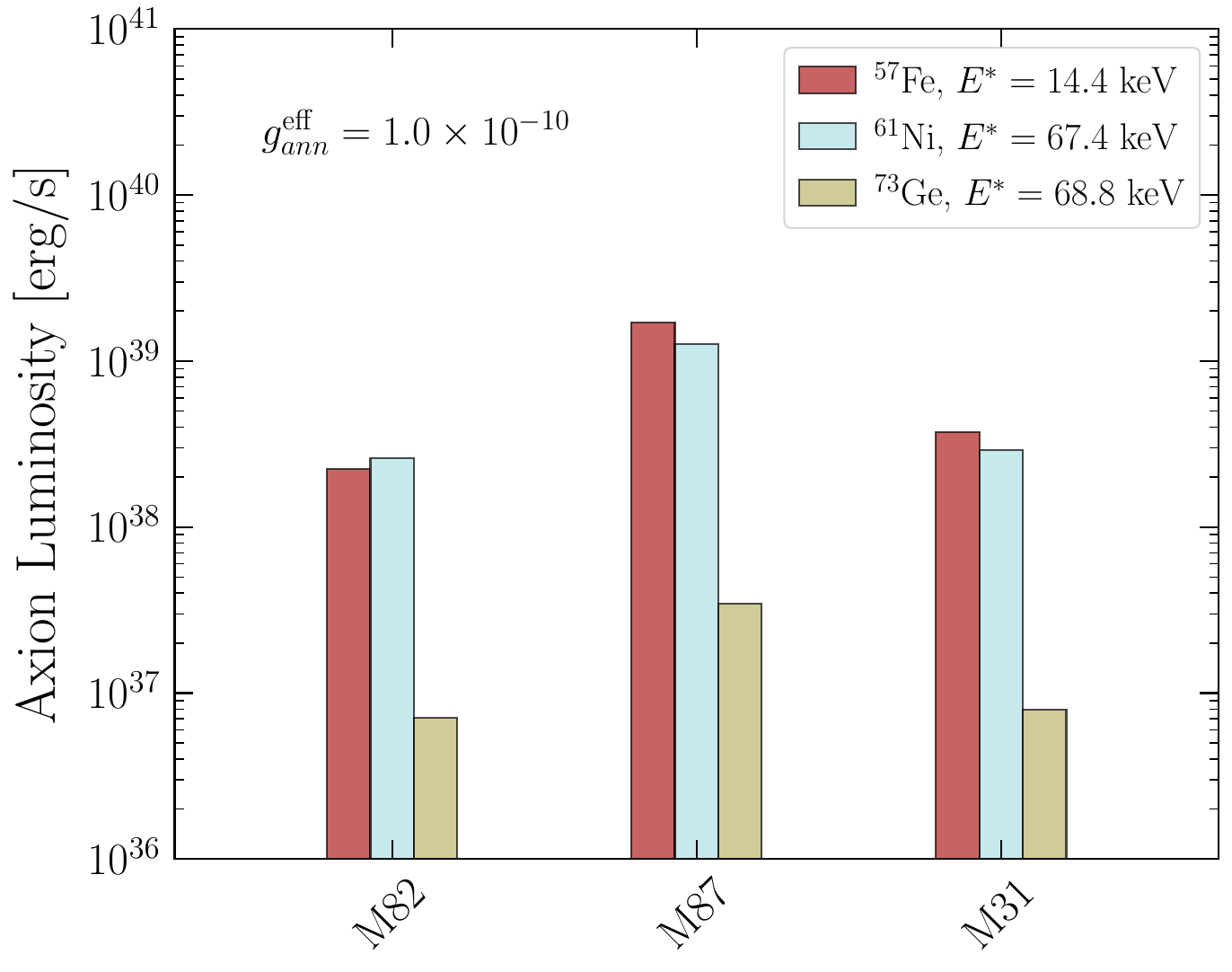}
\vspace{-0.4cm}
\caption{Comparison of the total axion luminosity from the indicated nuclear transitions across our M82, M87, and M31 stellar populations. We note that since our GC model is similar to that of M31, with corresponding luminosities $\sim$5 orders of magnitude smaller, we do not illustrate it here. Note that these luminosities are constructed with the indicated effective axion-nucleon coupling.  This figure shows that the $\Fe$ and $\Ni$ de-excitations are the most significant sources of axions.}
\label{fig:pop_axion_lum}
\end{figure}

In Fig.~\ref{fig:pop_axion_lum}, we illustrate a comparison of the total axion luminosity from our three strongest targets: M82, M87, M31, and the three main nuclear isotopes we consider in our work. We see that at the level of the raw axion emission, the three galaxies are roughly comparable, although M87, due to its larger number of stars, has the highest axion luminosity across all three isotopes. Additionally, we note that amongst the three galaxies, M82 has the highest ratio of $\Ni$ to $\Fe$ emission, which comes from its starburst nature and larger fraction of massive stars with higher core temperatures. This benefits $\Ni$ line emission slightly more than $\Fe$. Finally, we note that, of course, Fig.~\ref{fig:pop_axion_lum} only represents the axion emission from each galaxy; other aspects of the signal model, such as the conversion probabilities and distances, are also integral to the final photon spectrum for which we search.

\section{Searching for Ultralight Axions}

If axions are ultralight, they can efficiently convert to X-ray photons in the magnetic fields of our galactic targets. This idea was first fully explored in~\cite{Ning:2024eky}, where, as a summary, full-scale galactic models of magnetic fields and free-electron densities were chosen from ensembles of candidate galaxies in the IllustrisTNG TNG50 and TNG300 simulations~\cite{Pillepich:2019bmb, Nelson:2019jkf}. These analogue galaxies were filtered to have similar properties to the galaxies of interest, including star formation rate, stellar mass, and disk morphology. The associated magnetic field and free-electron density models allow for the calculation of the probability of an axion-to-photon conversion, $P_{a\to\gamma}$, given an axion mass $m_a$ (see, \textit{e.g.},~\cite{Ning:2024eky}).

\begin{figure}[!t]
\centering
\includegraphics[width=0.49\textwidth]{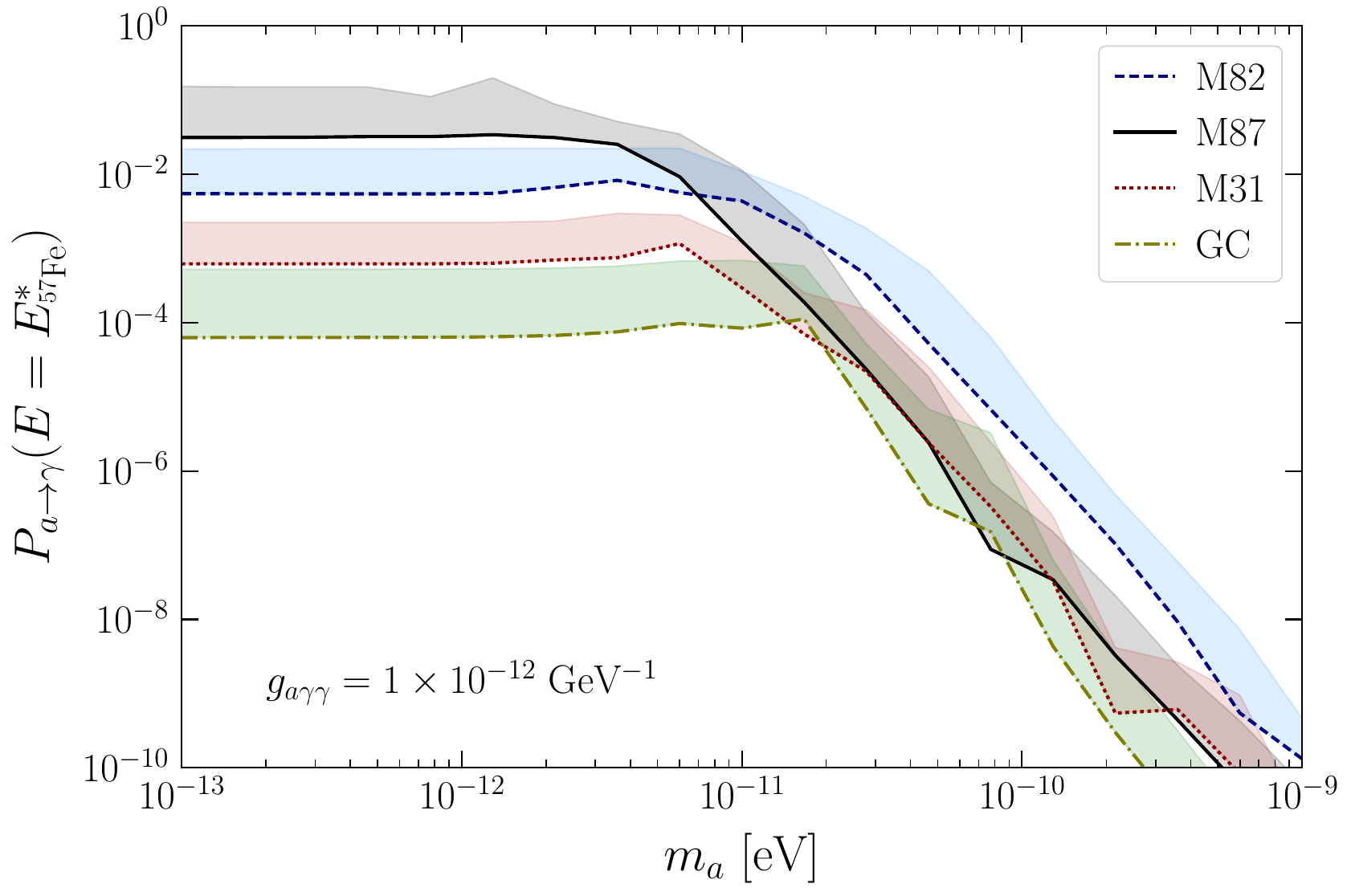}
\vspace{-0.4cm}
\caption{An illustration of the axion-to-photon conversion probabilities across the four galactic targets considered in this work. Here we show the conversion probabilities as a function of axion mass $m_a$, evaluated at the $\Fe$ line, \textit{i.e.} $E^* = 14.4$ keV. For M82, M87, and M31, whose conversion probabilities are calculated from IllustrisTNG models (see main text and~\cite{Ning:2024eky}), we show our fiducial conversion probabilities (bold lines) as well as the $1\sigma$ containment bands for the conversion probabilities across our galaxy candidates and orientations. For the Milky Way GC, we calculate conversion probabilities from parametric magnetic field models (see main text), and show the resulting containment bands, with the adopted probabilities given as the bold dash-dot line.}
\label{fig:conv_prob_ma}
\end{figure}

In this work we follow the same setup~\cite{Ning:2024eky} for modeling axion-to-photon conversion and refer to that work for further details, although we give a brief summary here. After selection of the analogue galaxy candidates in the simulations and randomly choosing orientations for a given galaxy, we draw stellar positions from the simulated baryonic distributions, and axion-to-photon conversion probabilities are calculated along the line-of-sight from those points. These conversion probabilities, derived from the ensemble of galaxy candidates and orientations, are then used to compute upper limits on the axion couplings.
Following~\cite{Ning:2024eky}, our choice of fiducial model is the conservative one that minimizes the expected sensitivity ({\it i.e.}, yields the weakest low-mass limit at 1$\sigma$ containment). In Fig.~\ref{fig:conv_prob_ma}, we give an illustration of our axion-to-photon conversion probabilities as a function of $m_a$, across our galaxy targets, evaluated at the $\Fe$ line energy. We also show the $1\sigma$ containment bands on these probabilities, derived from our usage of IllustrisTNG galactic magnetic field models of M82, M87, and M31. For the GC, we calculate conversion probabilities for the 8 parametric galactic magnetic field models in~\cite{Unger:2023lob} as well as the older model in~\cite{2012ApJ...757...14J}, and also assume the \texttt{ne2001} free-electron density model~\cite{Cordes:2002wz}. To account for uncertainties in the GC magnetic field modeling, we choose the weakest (in terms of conversion probability) of the parametric models detailed in~\cite{Unger:2023lob, 2012ApJ...757...14J} as our fiducial model, which is illustrated, along with the uncertainty bands, in Fig.~\ref{fig:conv_prob_ma}.  Note that for the case of the IllustrisTNG-based models, we show the uncertainty bands over the full ensemble of analogue galaxies and orientations, with details for how these analogues are chosen given in~\cite{Ning:2024eky, Ning:2025tit}. 

We emphasize an overall remark here that by construction our choices for the adopted fiducial conversion probabilities are meant to be conservative, as can be seen in Fig.~\ref{fig:conv_prob_ma}. This is especially important, since the magnetic field uncertainties dominate the uncertainties in our overall analysis, exactly analogous to the situations in~\cite{Ning:2024eky} and~\cite{Ning:2025tit} (we defer further discussion of the astrophysical population uncertainties to those works, as they are identical). Taking the more conservative side of the magnetic field modeling, as well as other previously discussed choices in our analysis such as the more conservative set of nuclear factors in our axion production calculation, gives us confidence that our final results are reasonably robust to systematic uncertainties. For example, taking instead the median conversion probabilities instead of our fiducial, lower $1\sigma$ values for M82, M87, M31, and the GC would result in our low-$m_a$ upper limits on $\ganneffgagg$ being 1.4, 1.5, 1.4, and 1.7 times stronger, respectively, than our fiducial limits.

Proceeding forward, these conversion probabilities are finally combined with our axion luminosity signal to construct the expected axion-induced photon spectrum at Earth for a given galaxy distance $d$:
\begin{equation}
    \Phi = L_a \times P_{a\to\gamma}(m_a,E^*) \times {1 \over 4\pi d^2} \,.
\end{equation}
Above, we explicitly emphasize the energy and axion mass dependence of the conversion probability. In particular, the axion conversion probability becomes suppressed when, roughly, the axion and photon become out of phase over the coherence length of the magnetic fields, which roughly translates to $\delta k L \sim {m_a^2 \over 2 E^*} {\rm kpc} > 1$, with $\delta k$ the momentum mismatch between the axion and photon and $L \sim {\rm kpc}$ the typical galactic magnetic field coherence length. (Note that in practice we exactly solve the axion-photon mixing equations at the relevant $E^*$ and $m_a$.)
We forward model this monochromatic flux through the NuSTAR detector, to find a line-like signal that is spectrally broadened at the level of $\sim$$10\%$.

We search for the resulting axion-induced spectra using NuSTAR data towards the appropriate targets, following the procedures and analyses developed in~\cite{Ning:2024eky, Ning:2025tit} with the crucial difference being that the signal considered here is not spectrally broad but is instead a narrow line at a known energy $E^*$. First, we compile archival NuSTAR data for M82, M87, M31, and the GC, for each of the two NuSTAR Focal Plane Modules (FPM). Our archival data for M82 and M87 are detailed in~\cite{Ning:2024eky}, while that for M31 is detailed in~\cite{Ning:2025tit}, and that for the GC is listed in Table~\ref{tab:GC_obs} in App.~\ref{app:figures}. We reduce the data using HEASoft version 6.28~\cite{2014ascl.soft08004N}, ultimately deriving counts spectra for source (ON) and background (OFF) regions, as defined in~\cite{Ning:2024eky, Ning:2025tit}, as well as the redistribution matrix files (RMFs) and auxiliary response files (ARFs) used for forward modeling the axion signal.   
Unlike the methods in~\cite{Ning:2024eky, Ning:2025tit}, we keep the counts spectra in bins defined by the native NuSTAR energy resolution channels (with bin widths $\sim$\,40 eV) since we are searching for a narrow spectral feature.

The forward-modeled axion signal is computed by
\begin{equation}
    \mu^{\rm tot}_{S, i}(\boldsymbol{\theta}_S) = \sum_et^e \int dE' \text{RMF}_i^e(E') \text{ARF}^e(E') S(E' | \boldsymbol{\theta}_S).
\label{eq:mu_S}
\end{equation} 
where we sum over the different exposures with exposure times $t^e$. The physical axion spectrum $S$ has units of [cts/cm$^2$/s/keV] and has support only in a single input energy bin, with signal parameters $\boldsymbol{\theta}_S = \{ m_a,\ganneffgagg$\}. The quantity $\mu^{\rm tot}_{S,i}$ reflects the expected signal counts for signal $S$ in energy bin $i$.  We combine the signal model with a power-law background model, with two nuisance parameters $\boldsymbol{\theta}_B = \{ \alpha, \beta \}$ for the normalization and spectral index, to describe the data under the null hypothesis. This combined signal and astrophysical background is then compared to the NuSTAR OFF-subtracted counts data using a Gaussian likelihood. We profile over the background nuisance parameters using standard frequentist techniques~\cite{Cowan:2010js, Cowan:2011an,Safdi:2022xkm} to obtain, given a fixed $m_a$, quantities such as the best-fit signal, the 95\% upper limit on our signal parameters $\ganneffgagg$, as well as the $1\sigma$ and $2\sigma$ expectations for the 95\% upper limits under the null hypothesis.

Note that in performing the analyses we select data within $\pm 3 \sigma$ of the line energy $E^*$, with $\sigma$ defined as the approximate standard deviation of the NuSTAR energy resolution at central energy $E^*$. For $\Fe$ ($\Ni$) ($\Ge$) we find $\sigma \approx 0.19$ keV ($\sigma \approx 0.29 $ keV) ($\sigma \approx 0.29$ keV). We also note that our analysis energy ranges do not coincide with known NuSTAR instrumental lines~\cite{Wik:2014}. 

\begin{figure}[!t]
\centering
\includegraphics[width=0.49\textwidth]{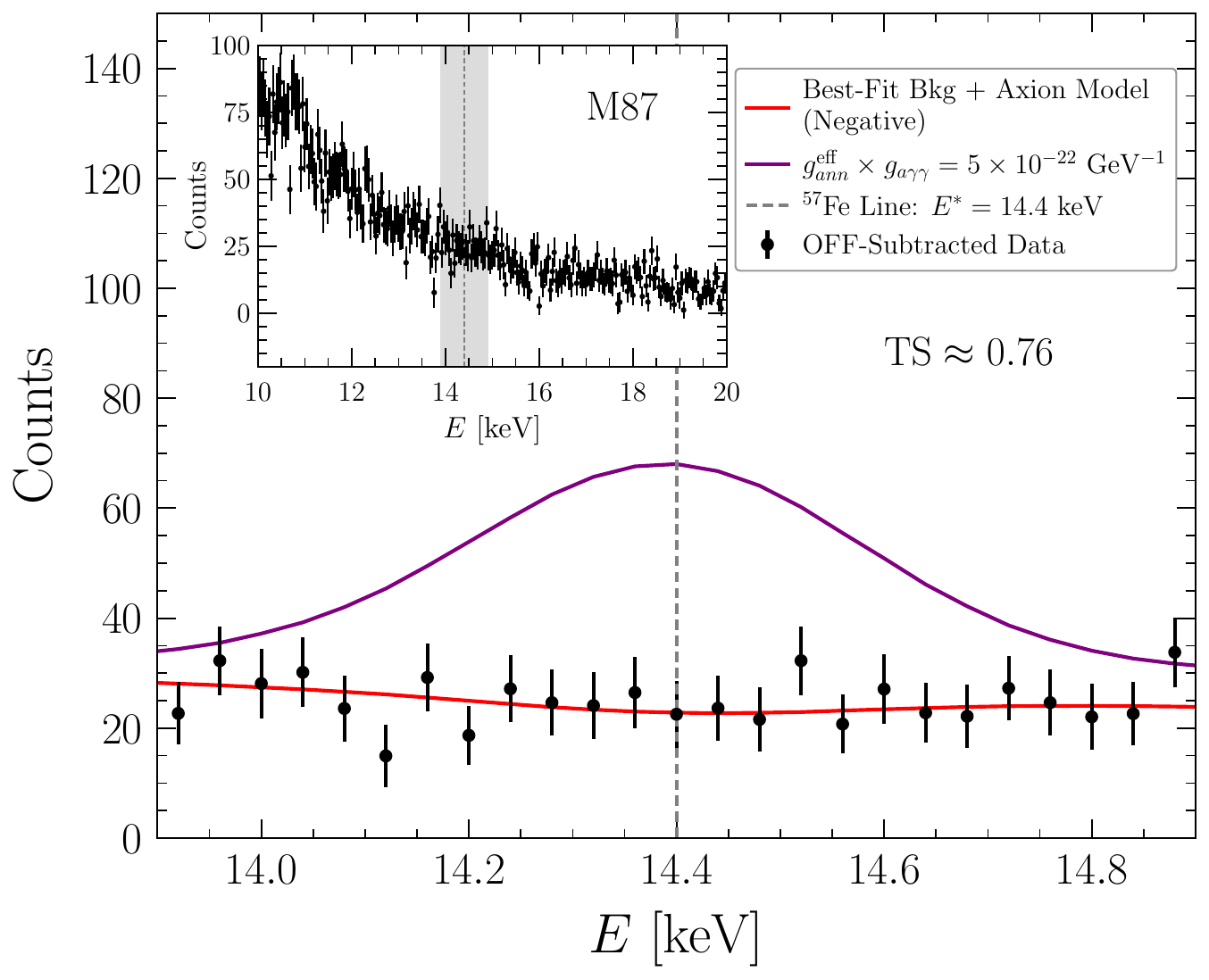}
\vspace{-0.4cm}
\caption{Illustration of the data analysis performed for the $\Fe$ line over the stellar population of M87. The inset shows the broader region around the $E^* = 14.4$ keV line, with the main plot showing the best-fit axion signal and background model (red), compared against the OFF-subtracted NuSTAR data in the analysis energy range around the expected line emission. Also shown is a large example axion-induced signal (purple) with the indicated couplings. The signals shown here are taken in the low $m_a$ limit.}
\label{fig:example_data_analysis}
\end{figure}

In Fig.~\ref{fig:example_data_analysis} we illustrate the best-fit signal model and background over our OFF-subtracted NuSTAR data for our most stringent constraint, the search for $\Fe$ from M87. The signal shows up as a narrow bump in the spectra around the de-excitation energy $E^*=14.4$ keV.  More illustrations, \textit{i.e.} those for a selection of our other targets, can be found in App.~\ref{app:figures}.

Ultimately, we find no evidence for axions in our targets and for the nuclear isotope de-excitation energies considered (see Tab.~\ref{tab:results}). Interpreting our results first under the fiducial assumption of $\gann = \gapp$, we show our two most stringent 95\% upper limits and their $1/2\sigma$ containment intervals on $\ganngagg$ under the null hypothesis in Fig.~\ref{fig:ganngagg} (a more complete illustration of our results can be found in App.~\ref{app:figures}). Our most stringent constraints come from $\Fe$ searches from M87 and M82, where we constrain $|\ganngagg| \lesssim 1.1 \times 10^{-22}$ GeV$^{-1}$, and $|\ganngagg| \lesssim 3.1 \times 10^{-22}$ GeV$^{-1}$, respectively, with the significance of axion discovery being $0.87\sigma$ and $0.17\sigma$, respectively. We also derive slightly less competitive constraints from the $\Fe$ search in M31 and the GC, as well as the $\Ni$ searches from M82 and M87 (see App.~\ref{app:figures}). 

\begin{table*}[!htb]
\begin{tabular}[t]{p{1.5cm}p{1.5cm}p{1.5cm}p{2.5cm}p{2.5cm}p{2.5cm}}
Isotope & Energy [keV] & Target & discovery TS & $\ganneffgagg$ \newline [GeV$^{-1}$] \newline 95\% U.L.  & $\ganneffgagg$ \newline [GeV$^{-1}$] \newline B.F.\\ \hline \hline
$\Fe$       & 14.4 & M82 & 0.03 & $4.0 \times 10^{-22}$ & $7.6 \times 10^{-23} $      
\\
       &  & M87 & 0.76 & $1.4 \times 10^{-22}$ & $(-)$       
\\
       &  & M31 & 0.30 & $5.6 \times 10^{-22}$ & $(-)$      
\\
       &  & GC & 0.17 & $1.2 \times 10^{-21}$ & $(-)$       
\\
\hline
$\Ni$       & 67.4 & M82 & 1.5 & $1.2 \times 10^{-21}$ &  $7.9 \times 10^{-22}$      
\\
       &  & M87 & 1.9 & $1.0 \times 10^{-21}$ & $6.8 \times 10^{-22}$       
\\
       &  & M31 & 1.9 & $4.2 \times 10^{-21}$ & $3.0 \times 10^{-21}$       
\\
       &  & GC & 0.68 & $2.3 \times 10^{-21}$ & $(-)$       
\\
\hline
$\Ge$       & 68.8 & M82 & 0.18 & $4.7 \times 10^{-21}$ & $(-)$       
\\
       &  & M87 & 2.5 & $5.3 \times 10^{-21}$ & $3.5 \times 10{-21}$       
\\
       &  & M31 & 0.06 & $1.9 \times 10^{-20}$ & $5.3 \times 10^{-21}$       
\\
       &  & GC & 1.4 & $1.1 \times 10^{-20}$ & $(-)$       
\\
\hline 
\hline
\end{tabular}
\caption{\label{tab:results} The resulting discovery TS's for a two-sided test, 95\% upper limits (U.L.) and best fits (B.F.) for $\ganneffgagg$ from our search for nuclear de-excitation-induced axions over our selected galaxy targets and nuclear isotopes. The upper limits and best fits are quoted for the massless axion limit, and the $(-)$ symbol indicates a negative best fit. Finding no significant evidence for axions, our most stringent constraints come from $\Fe$ searches in M82 and M87.} 
\end{table*}

\section{Discussion}

In this work we derive leading constraints on $|g_{ann}^{\rm eff} \times g_{a\gamma\gamma}|$ for ultralight axions with masses $m_a \lesssim 10^{-10}$ eV, where $g_{ann}^{\rm eff}$ is approximately equal to $g_{ann}$ but includes a small admixture of $g_{app}$.  Axions coupled to nucleons through $g_{ann}^{\rm eff}$ may be emitted efficiently from nuclear de-excitations in massive, hot stars with core temperatures near the excitation energies $E^*$. Our strongest constraints come from considering the $\Fe$ transition, though $\Ni$ and $\Ge$ are competitive. We consider axions emitted from nearby stellar populations in the Milky Way, M31, M82, and M87.  While the emission rates are not sensitive to $m_a$ for $m_a \lesssim {\rm keV}$, the efficient conversion of these axions to X-rays in the magnetic fields permeating these galaxies requires ultralight axion masses. 
 We search for the converted hard X-rays using data from NuSTAR towards these targets.  Our strongest constraints arise from M87 partially due to the efficient axion-to-photon conversion rates in the Virgo cluster that hosts M87. 

As we show in Fig.~\ref{fig:ganngagg}, our results disfavor the axion interpretation of the M7 neutron star X-ray excess~\cite{Buschmann:2019pfp} for axion masses below roughly $10^{-10}$ eV by over two orders of magnitude in $|g_{ann} \times g_{a\gamma\gamma}|$.  In that figure we show the $1\sigma$ best-fit parameter space to explain the excess high-energy X-rays observed towards the nearby M7 neutron stars by the Chandra and XMM-Newton telescopes~\cite{Dessert:2019dos}. These X-ray excesses could be explained by axions produced within the cores of these neutron stars through nuclear bremsstrahlung and then subsequently the axions converting to X-rays in the neutron star magnetic fields.  However, our work shows that if the M7 excess is indeed arising from axions, the axion mass must be above roughly $10^{-10}$ eV. 

While this work requires the axion to couple to both nucleons and photons, it is worth contextualizing our results with those of searches that are only sensitive to the axion-photon coupling. In Fig.~\ref{fig:gagg_tree_loop} we show our results in the $|g_{a\gamma\gamma}|$  -- $m_a$ plane under two different assumptions for the dimensionless EFT coefficients $C_{ann} / C_{a\gamma\gamma}$.  Under the assumption $C_{ann} = C_{a\gamma\gamma}$, which would be expected in {\it e.g.} DFSZ type UV completions~\cite{DiLuzio:2020wdo}, our results surpass in sensitivity those of all prior searches for $g_{a\gamma\gamma}$ alone at low masses. Thus, in classes of models where axions couple strongly to quarks, our search is the most sensitive to date. On the other hand, in models such as KSVZ where the axion has no tree-level coupling to quarks and the axion-quark couplings are loop induced (see App.~\ref{app:axionnucleon}), our constraints are sub-leading relative to existing searches for $g_{a\gamma\gamma}$ alone. 

In this work we show that axions are produced in abundance through nuclear de-excitations in nearby galaxies.  It would be interesting to try to use this aspect of our analysis, independent of the axion-to-photon conversion phenomena which is crucial to this work but requires low axion masses, to probe higher mass axions. For example, heavier axions could decay in-flight (see, {\it e.g.},~\cite{Candon:2024eah}) or more strongly coupled axions could lead to such large energy losses in parts of the stellar population that the back-reaction of the axions on the stellar population modeling would be important. We leave such studies to future work.  

\begin{acknowledgements}
{
{\it
We thank Josh Benabou, Francisco Cand\'{o}n, Andrea Caputo, Chris Dessert, Josh Foster, Maurizio Giannotti, Wick Haxton, Mathieu Kaltschmidt, Giuseppe Lucente, and Yujin Park for helpful conversations and discussions. O.N. and B.R.S. are supported in part by the DOE award DESC0025293. B.R.S. acknowledges support from the Alfred P. Sloan Foundation.  The work of O.N. is supported in part by the NSF Graduate Research Fellowship Program under Grant DGE2146752.  A.R. acknowledges support from the National Science Foundation (Grant No. PHY-2020275), and the Heising-Simons Foundation (Grant 2017-228). This research used resources of the National Energy Research Scientific Computing Center (NERSC), a U.S. Department of Energy Office of Science User Facility located at Lawrence Berkeley National Laboratory, operated under Contract No. DE-AC02-05CH11231 using NERSC award HEP-ERCAP0023978.}
}
\end{acknowledgements}

\appendix

\section{Additional Figures and Tables}
\label{app:figures}

\begin{figure}[!t]
\centering
\includegraphics[width=0.49\textwidth]{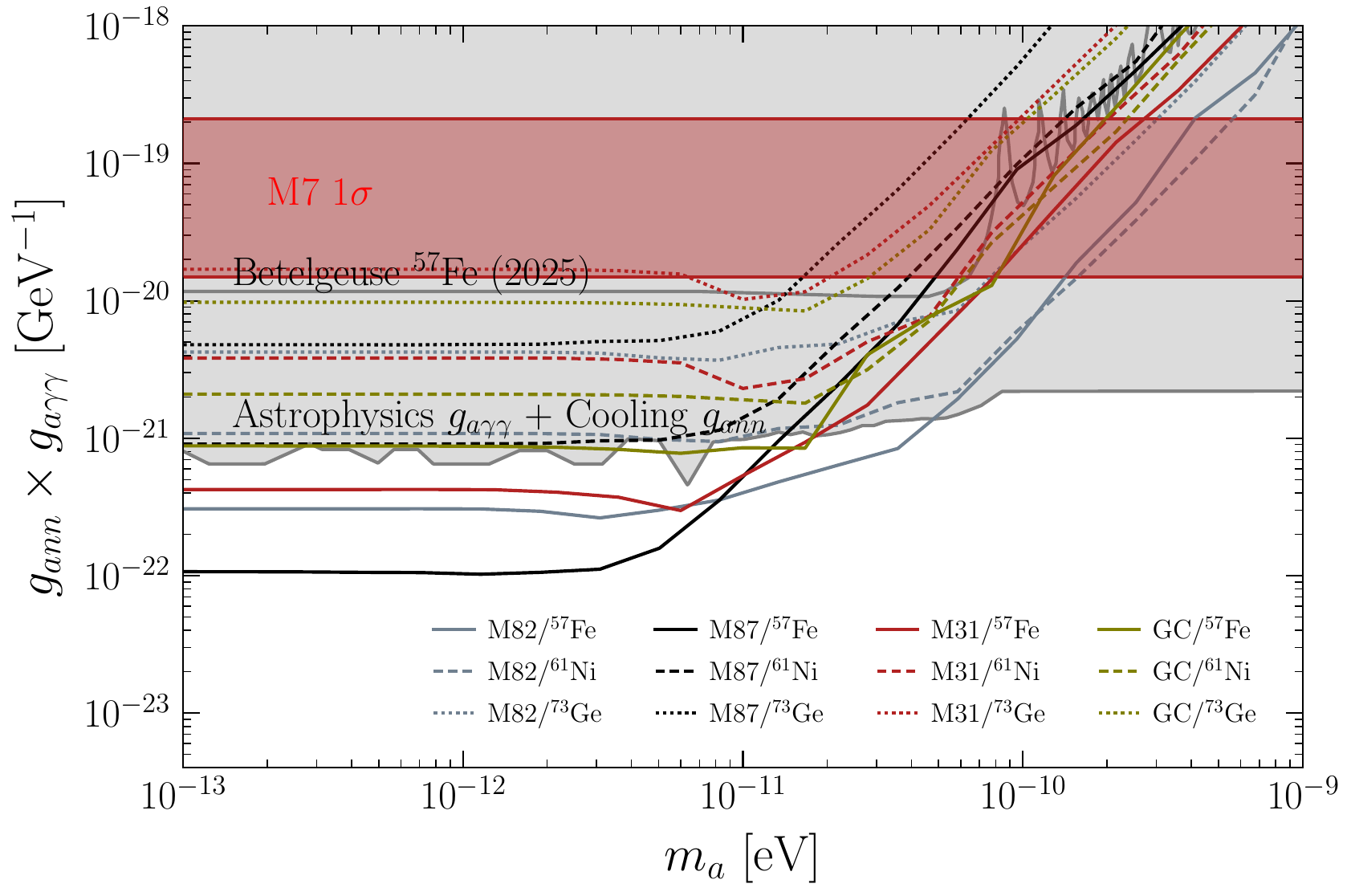}
\vspace{-0.4cm}
\caption{The 95\% upper limits on the combined axion-nucleon axion-photon coupling $\ganngagg$ from our search for ultralight axions produced from nuclear isotope de-excitation, across our four selected galactic targets and our three selected nuclear isotopes. In gray we illustrate existing astrophysical constraints, with more details in the main text (the red band indicates the $1\sigma$ signal for axions found in~\cite{Buschmann:2019pfp}).}
\label{fig:ganngagg_full}
\end{figure}

\begin{figure}[!t]
\centering
\includegraphics[width=0.49\textwidth]{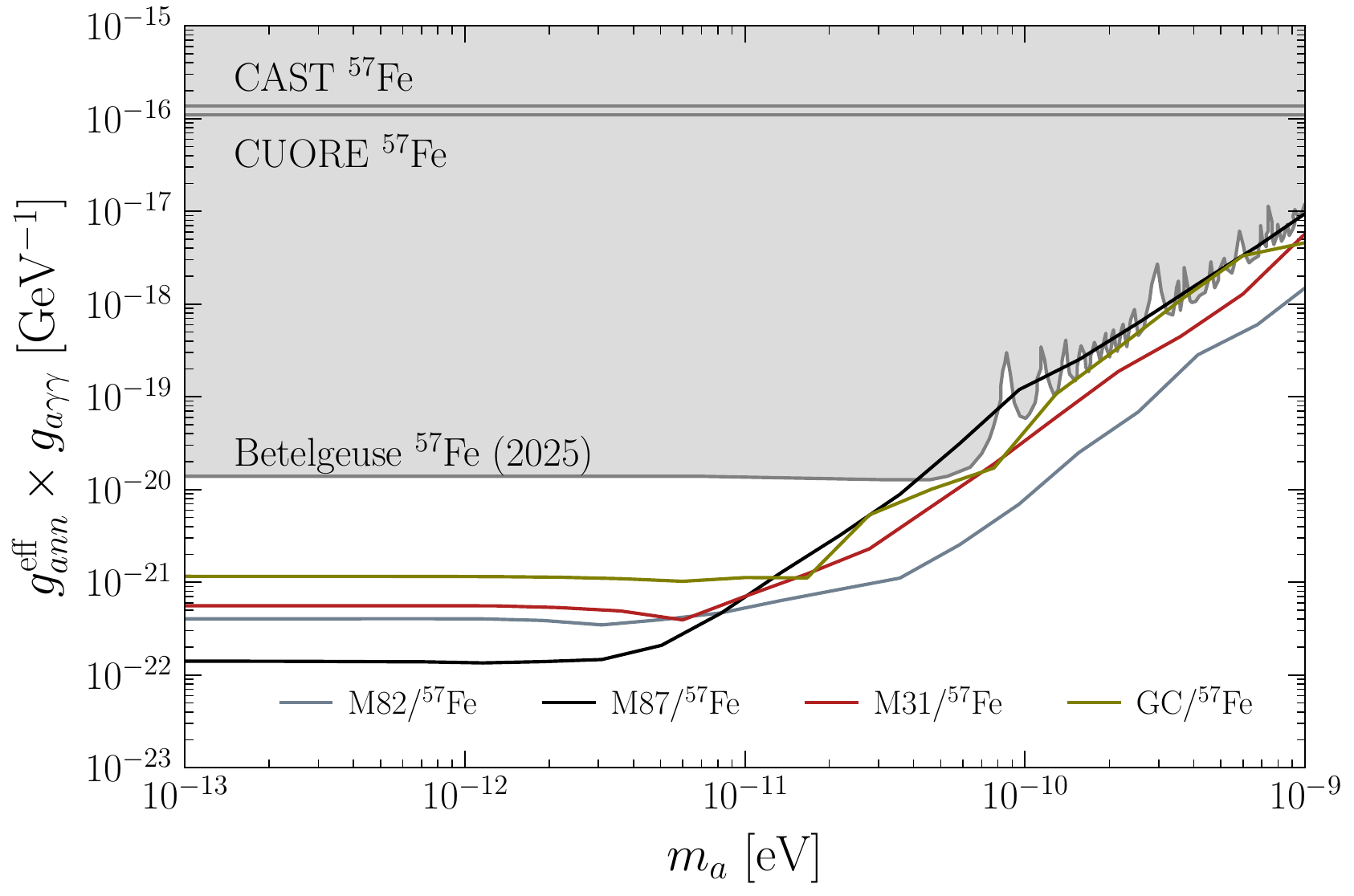}
\vspace{-0.4cm}
\caption{Similar to Fig.~\ref{fig:ganngagg_full}, although only illustrating constraints which directly probe $\ganneffgagg$, which include~\cite{CAST:2009jdc, Li:2015tyq, Candon:2025vpv} (rescaled using our $\Gamma_a/\Gamma_{\gamma}$ coefficient in~\eqref{eq:fe57_ganneff}). We show only our analogous constraints for $\Fe$ over our galactic targets, and display all results in terms of $\ganneffgagg$.}
\label{fig:geff}
\end{figure}

\begin{figure}[!t]
\centering
\includegraphics[width=0.49\textwidth]{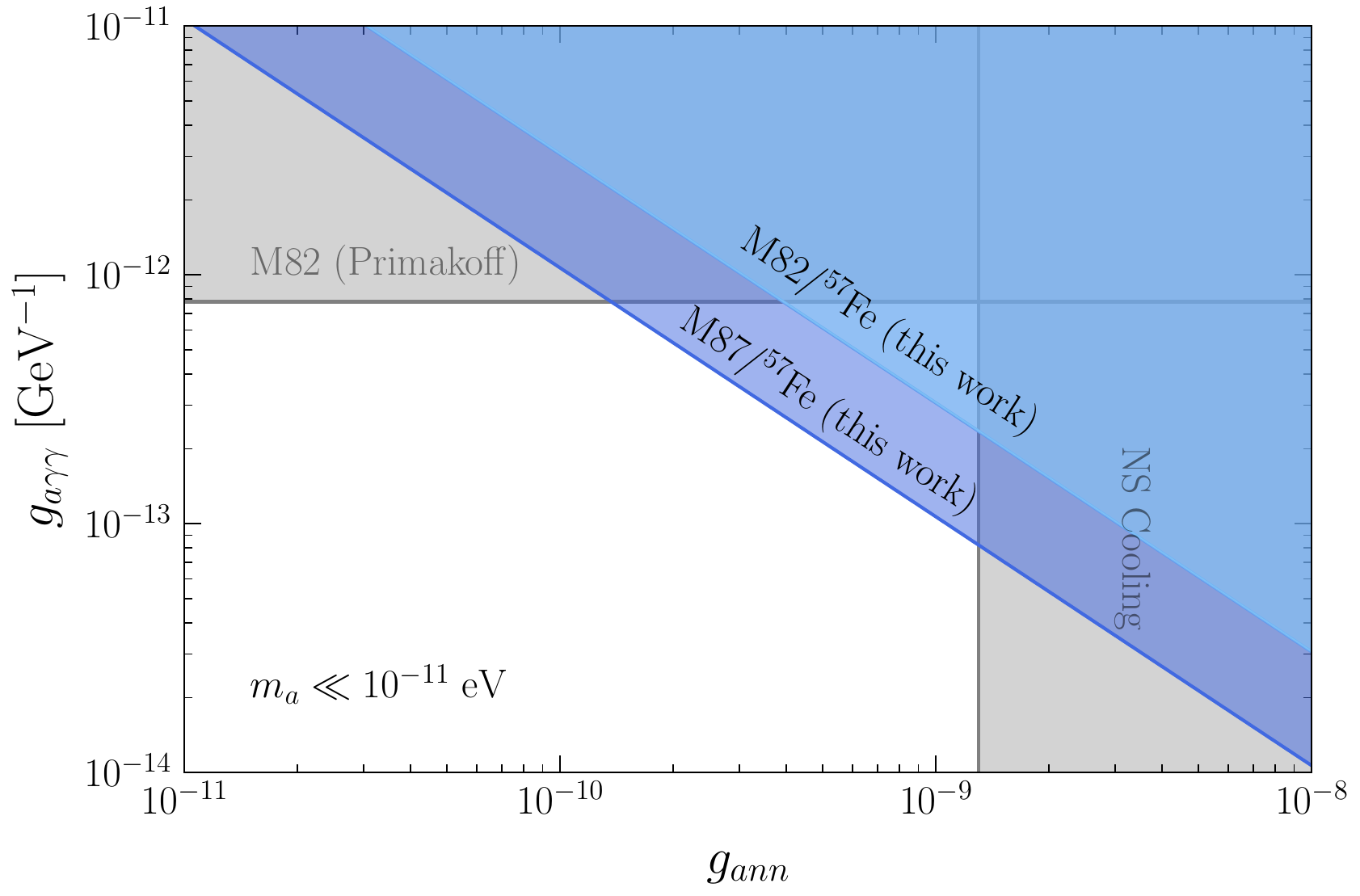}
\vspace{-0.4cm}
\caption{An illustration of our limits from Fig.~\ref{fig:ganngagg} recast as constraints on the $\gagg-\gann$ parameter space (for $\gann = \gapp$), in the massless axion limit. We illustrate the leading constraint in this limit on $\gagg$ from Primakoff production of axions from M82~\cite{Ning:2024eky}, as well as the leading constraint on $\gann$ from neutron star cooling~\cite{Buschmann:2021juv}.}
\label{fig:gann_vs_gagg}
\end{figure}

\begin{figure}[!t]
\centering
\includegraphics[width=0.49\textwidth]{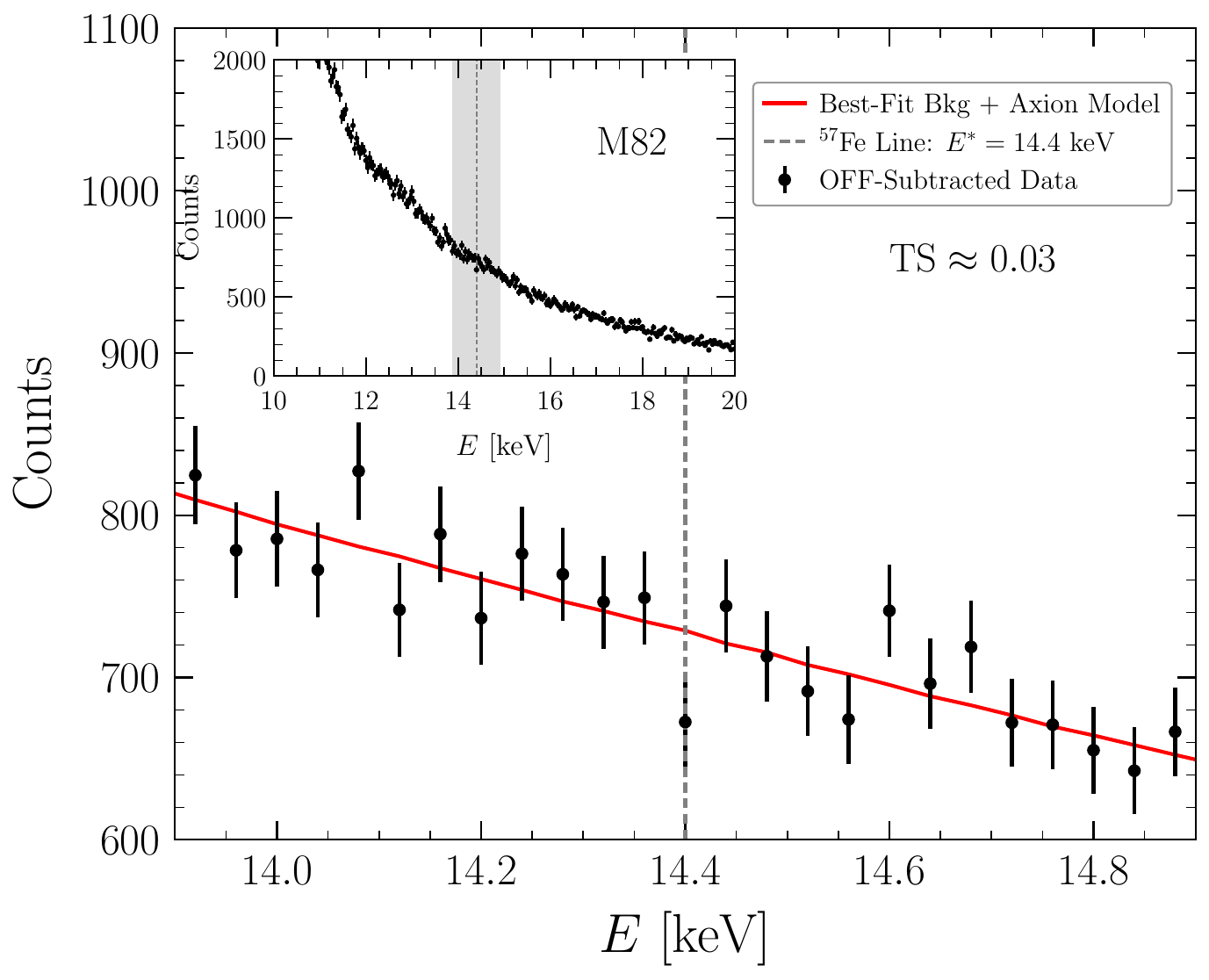}
\vspace{-0.4cm}
\caption{The same for Fig.~\ref{fig:example_data_analysis} but for M82, with the search for $\Fe$. Again, the inset shows the broader region around the $E^* = 14.4$ keV line, with the main plot showing the best-fit axion signal and background model, compared against the OFF-subtracted NuSTAR data in the $3\sigma$ region around the expected line emission.}
\label{fig:M82_example_data_Fe57}
\end{figure}

\begin{figure}[!t]
\centering
\includegraphics[width=0.49\textwidth]{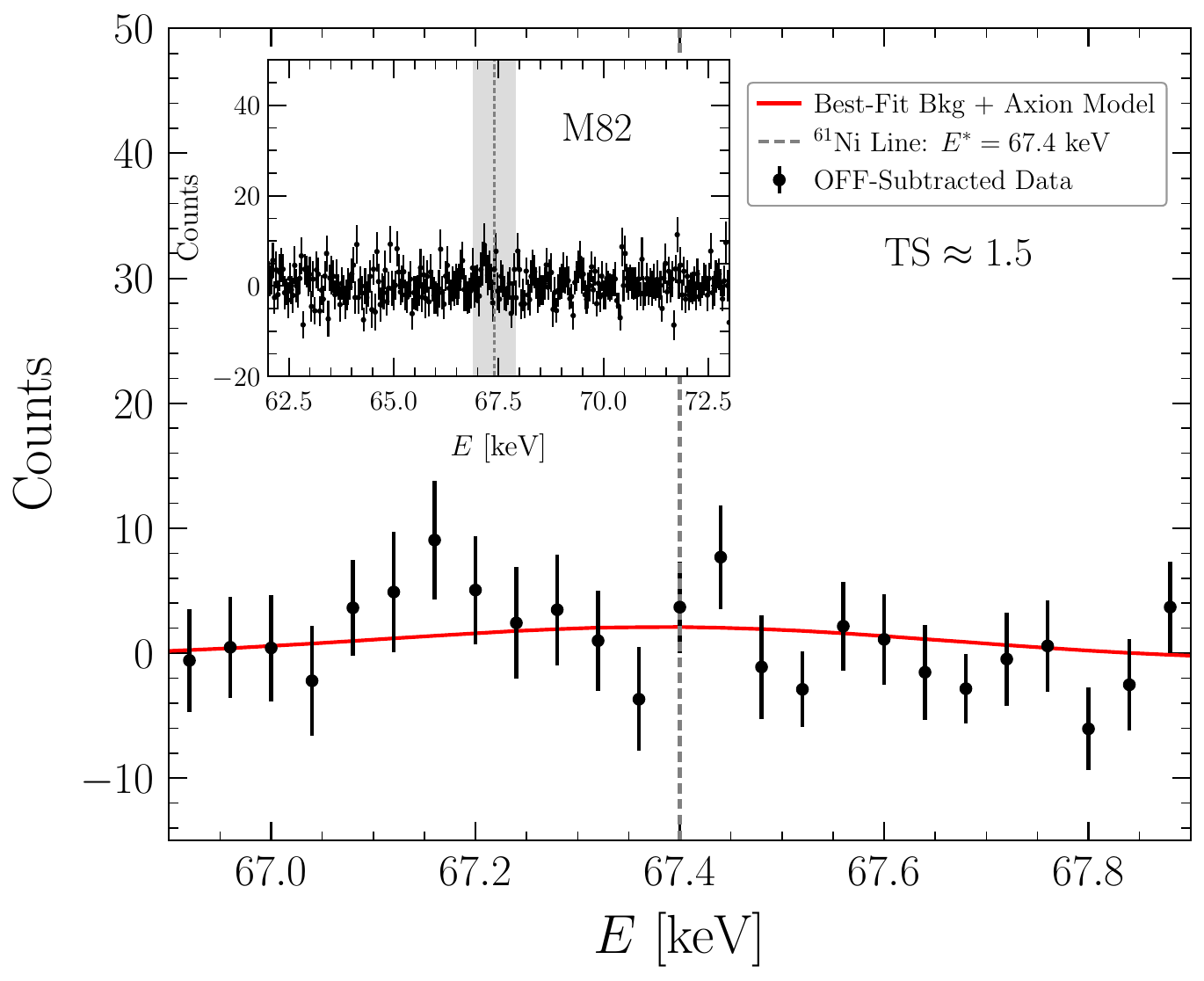}
\vspace{-0.4cm}
\caption{The same for Fig.~\ref{fig:example_data_analysis} but for M82, with the search for $\Ni$. The inset shows the broader region around the $E^* = 67.4$ keV line.}
\label{fig:M82_example_data_Ni61}
\end{figure}

\begin{figure}[!t]
\centering
\includegraphics[width=0.49\textwidth]{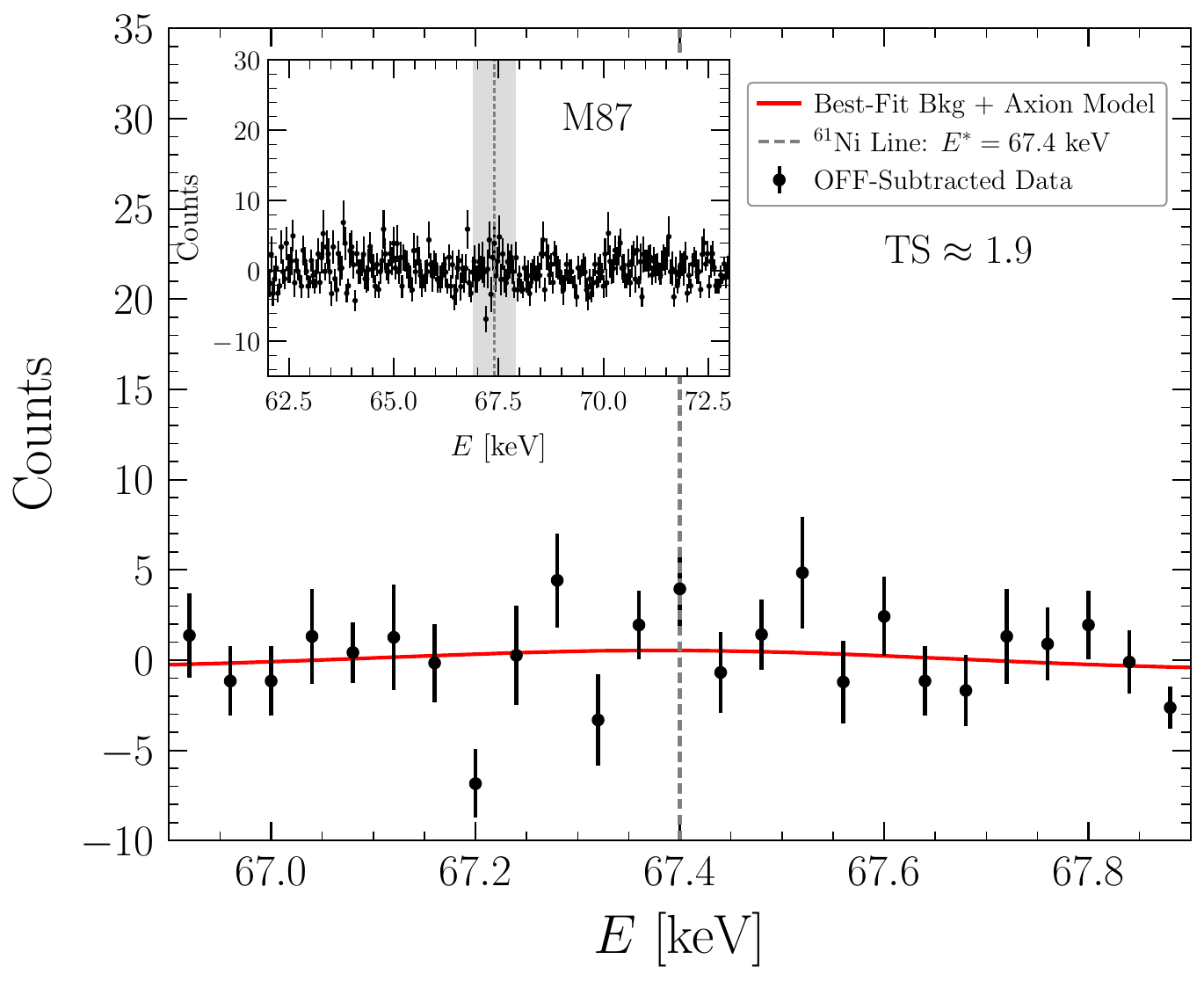}
\vspace{-0.4cm}
\caption{The same for Fig.~\ref{fig:example_data_analysis} but for M87, with the search for $\Ni$.}
\label{fig:M87_example_data_Ni61}
\end{figure}

\begin{figure*}[!htb]
\centering
\includegraphics[width=\columnwidth]{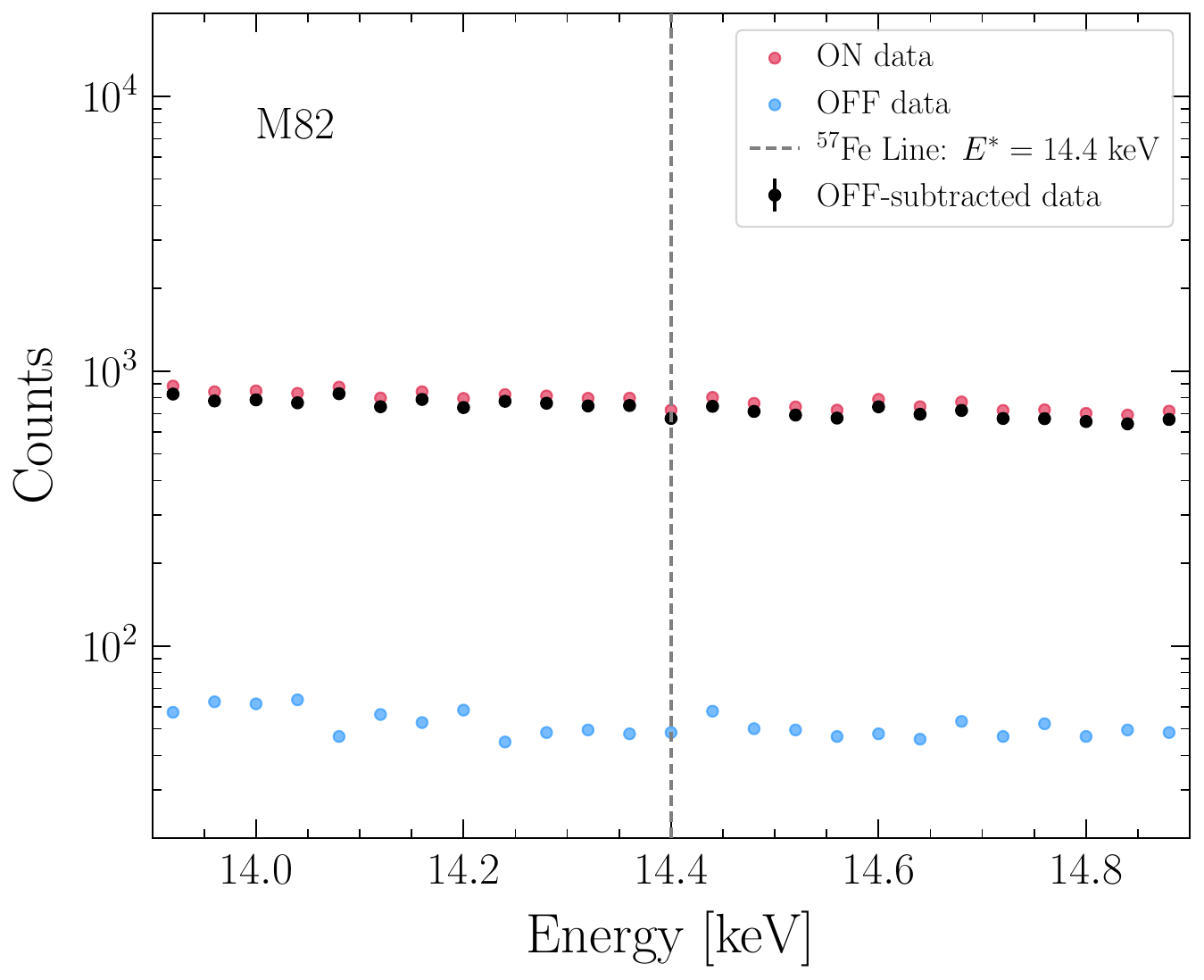}
\includegraphics[width=\columnwidth]{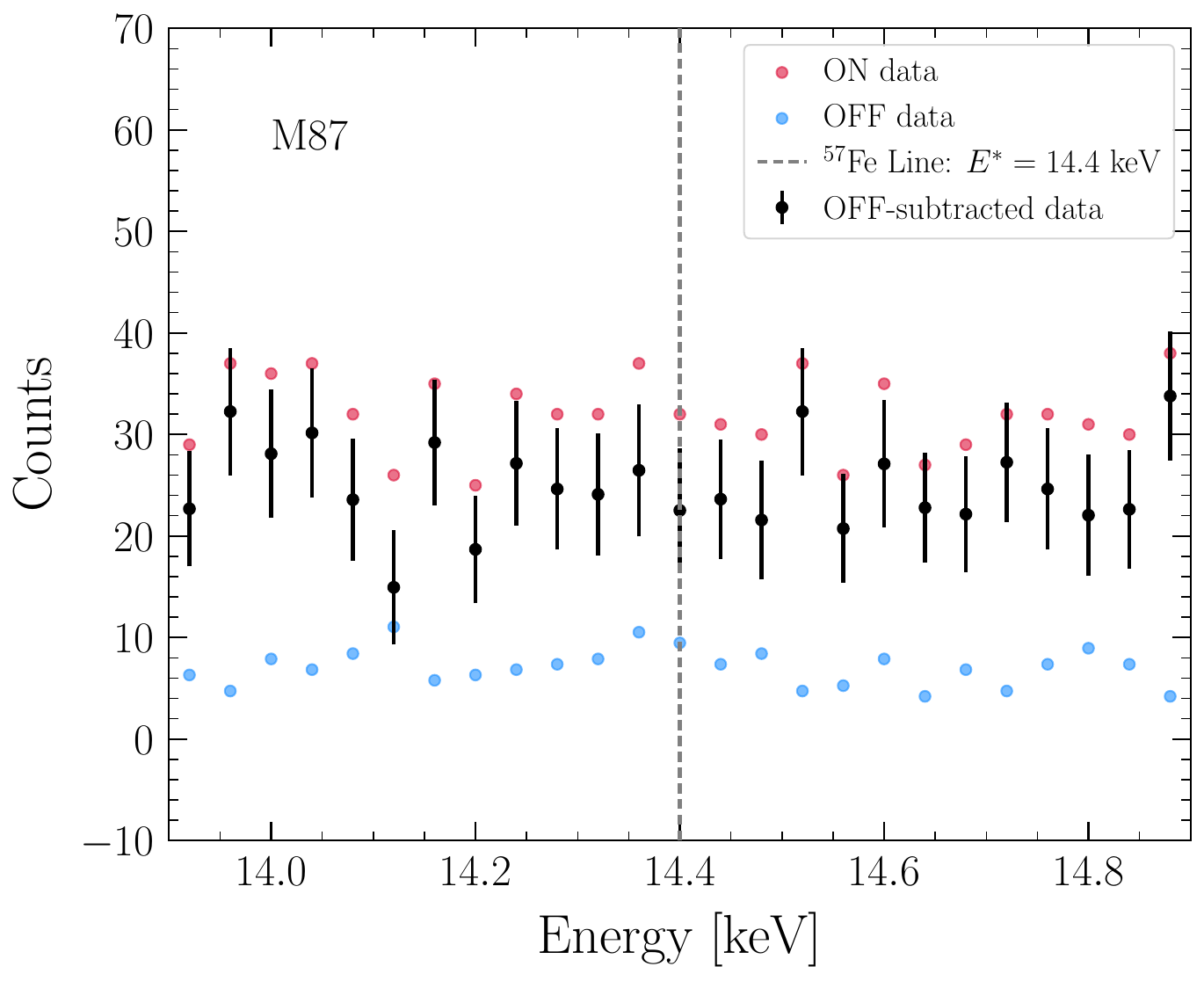}
\caption{(Left) An illustration of the ON data, the OFF data, and the OFF-subtracted data for our analysis of the $\Fe$ line search in M82, with 68\% Poisson uncertainties shown for the OFF-subtracted data. (Right) As in the left panel but for M87. We  see in both cases that the overall high number of counts per bin justifies the use of Gaussian errors in our analyses.}
\label{fig:more_data}
\end{figure*}

In this section we illustrate additional figures and a table closely related to those presented in the main text. In Fig.~\ref{fig:ganngagg_full} we illustrate the 95\% upper limits on $\ganngagg$ resulting from our complete searches across our four selected galactic targets and our three selected nuclear isotopes. In Fig.~\ref{fig:geff} we illustrate our $\Fe$ constraints cast as limits on the effective nucleon coupling $\ganneffgagg$, compared to other direct constraints on $\ganneffgagg$. Here we also rescale the constraint from~\cite{Candon:2025vpv} to our $\Gamma_a/\Gamma_{\gamma}$ coefficient in~\eqref{eq:fe57_ganneff}. Fig.~\ref{fig:gann_vs_gagg} is the constraint from Fig.~\ref{fig:ganngagg} recast as constraints on the $\gagg - \gann$ parameter space for $\gann = \gapp$ and massless axions. Figs.~\ref{fig:M82_example_data_Fe57},~\ref{fig:M82_example_data_Ni61} and~\ref{fig:M87_example_data_Ni61} are illustrations similar to Fig.~\ref{fig:example_data_analysis} where we show our data analyses searching for axions from $\Fe$ and $\Ni$ de-excitation lines across our two most powerful targets, M82 and M87. Also for these two targets and for $\Fe$ (the strongest constraints in this work), we illustrate in Fig.~\ref{fig:more_data} the explicit ON data, OFF data, and OFF-subtracted data used in our analyses. 

\begin{figure*}[!htb]
\centering
\includegraphics[width=\columnwidth]{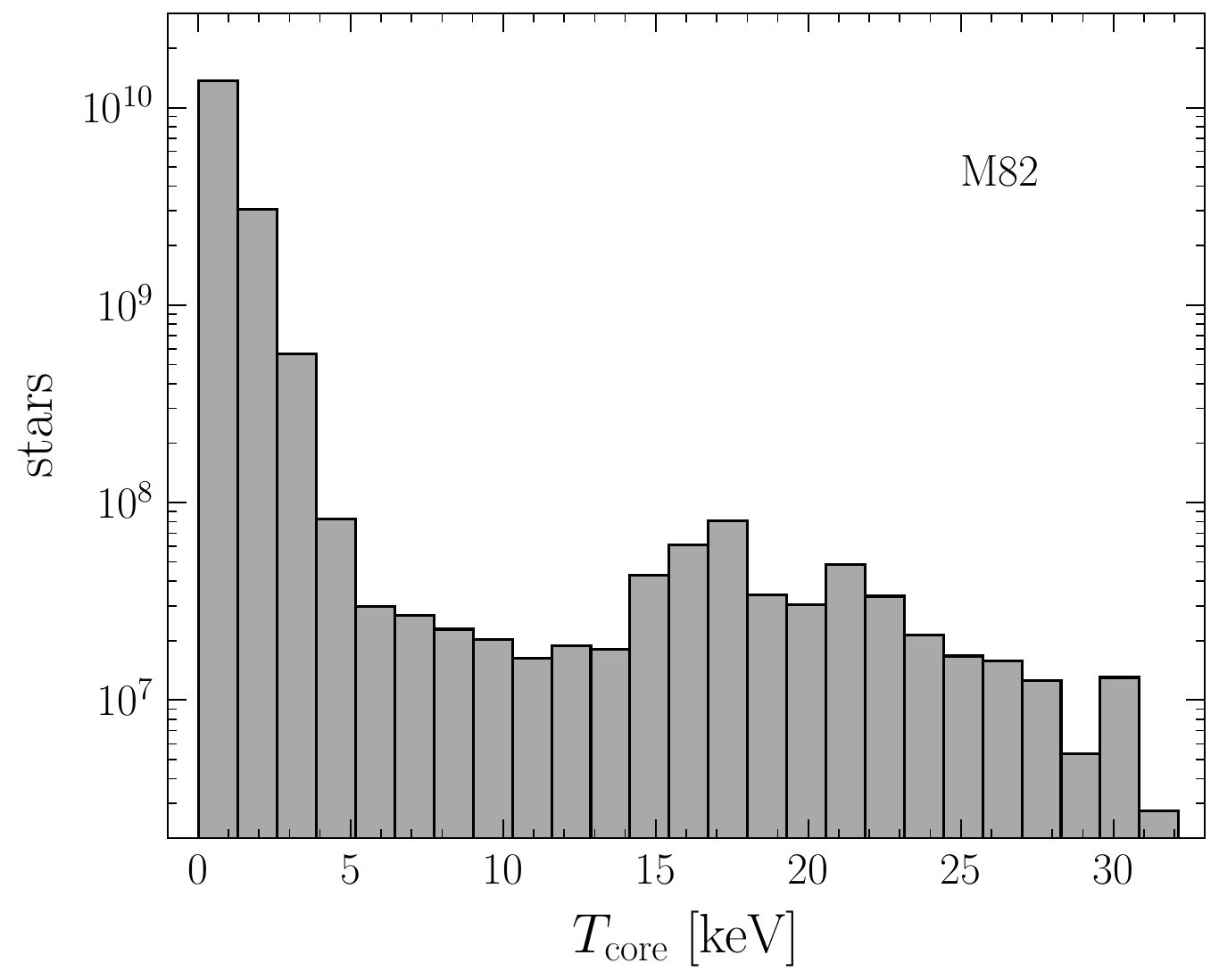}
\includegraphics[width=\columnwidth]{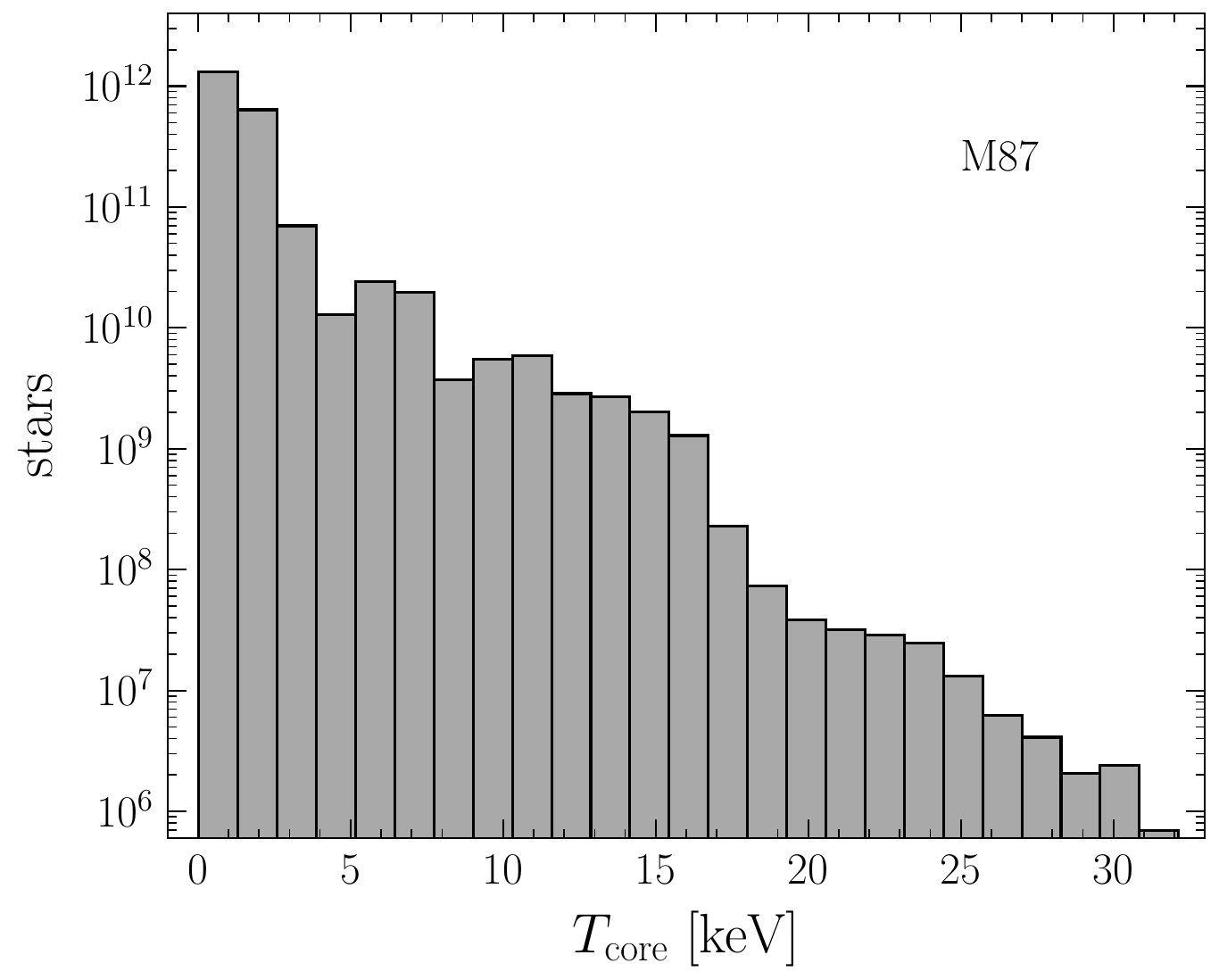}
\caption{(Left) The distribution of stellar core temperatures across our stellar population for M82. This illustration gives a broad idea of the available energies one has access to in the context of the search for nuclear line transitions discussed in this work. (Right) The same but for M87.}
\label{fig:stars_T}
\end{figure*}

\begin{figure}[!t]
\centering
\includegraphics[width=0.49\textwidth]{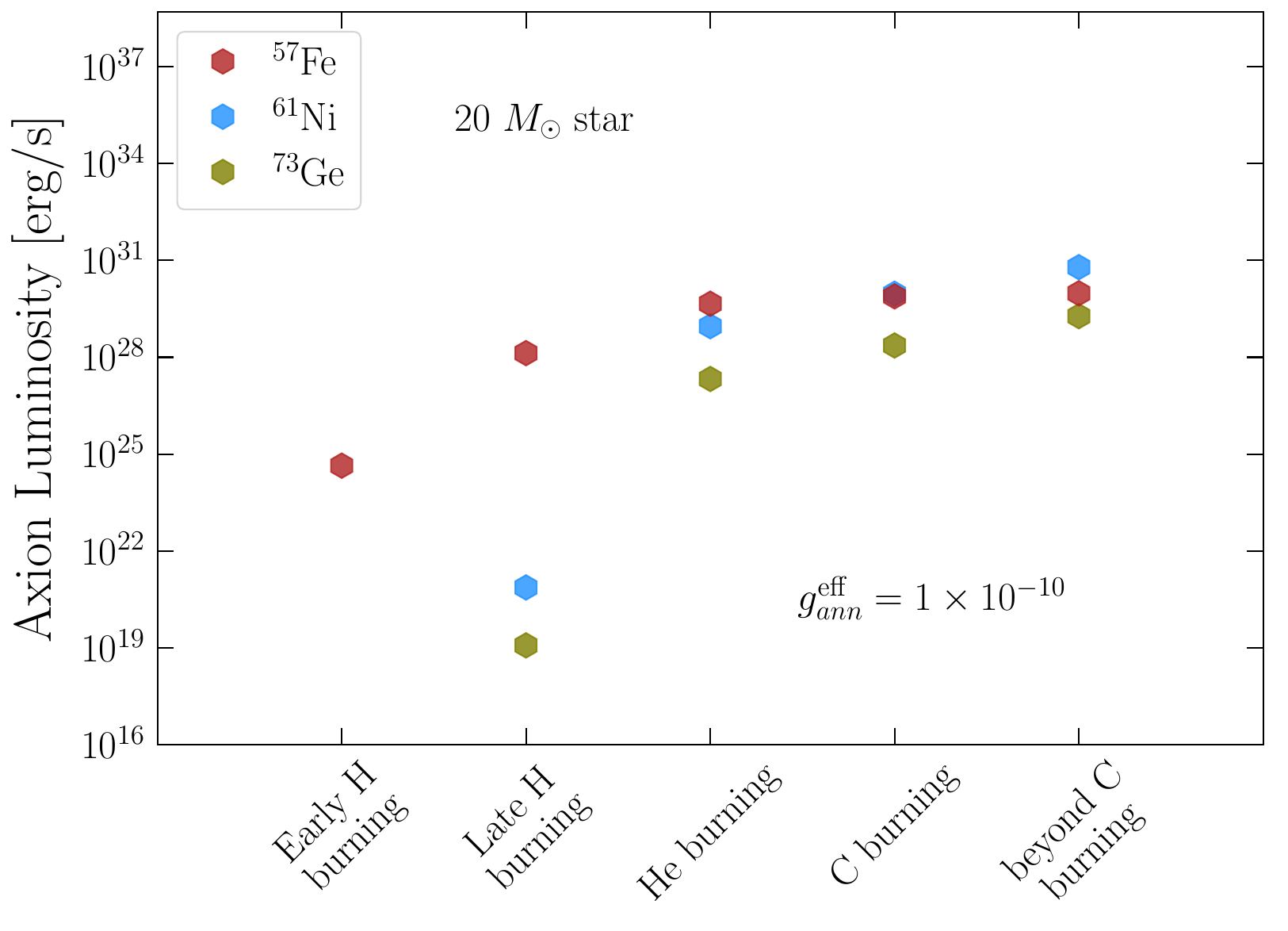}
\vspace{-0.4cm}
\caption{{An illustration of the total axion luminosity emitted from a 20 $M_{\odot}$ star, throughout various representative points in its lifetime. We show the luminosities for our three main isotopes at the indicated axion-nucleon coupling. We note that the axion luminosities are primarily due to primordial elemental abundances, given that heavy elements are not substantially produced in stellar interiors for the vast majority of a star's life (see main text).}}
\label{fig:evolution}
\end{figure}

\begin{figure}[!t]
\centering
\includegraphics[width=0.49\textwidth]{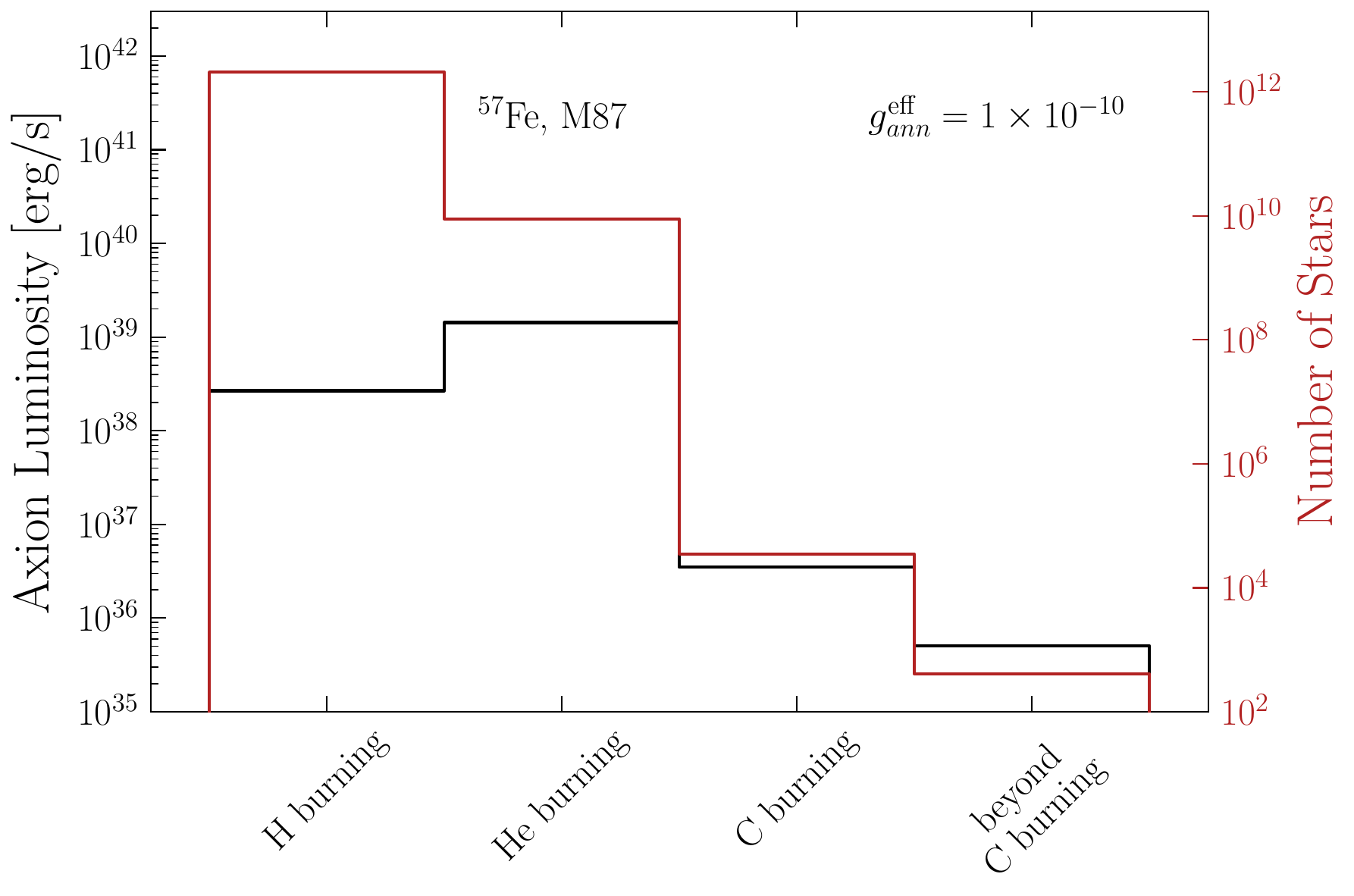}
\vspace{-0.4cm}
\caption{An illustration of the total axion luminosities (black) and the stellar numbers (red) across our stellar population in M87, distributed according to the main fusing element in the stellar core, for the $\Fe$ transition and with the indicated axion-nucleon coupling. Across our population, for our axion searches we are primarily sensitive to He core-burning stars. While there is evidence for a factor of a few enhancement in produced heavy elements such as $\Fe$ mostly during the C burning stages and beyond~\cite{Candon:2025vpv}, this would only have a minor effect on our overall population signal.}
\label{fig:el_burn}
\end{figure}

\begin{figure}[!t]
\centering
\includegraphics[width=0.49\textwidth]{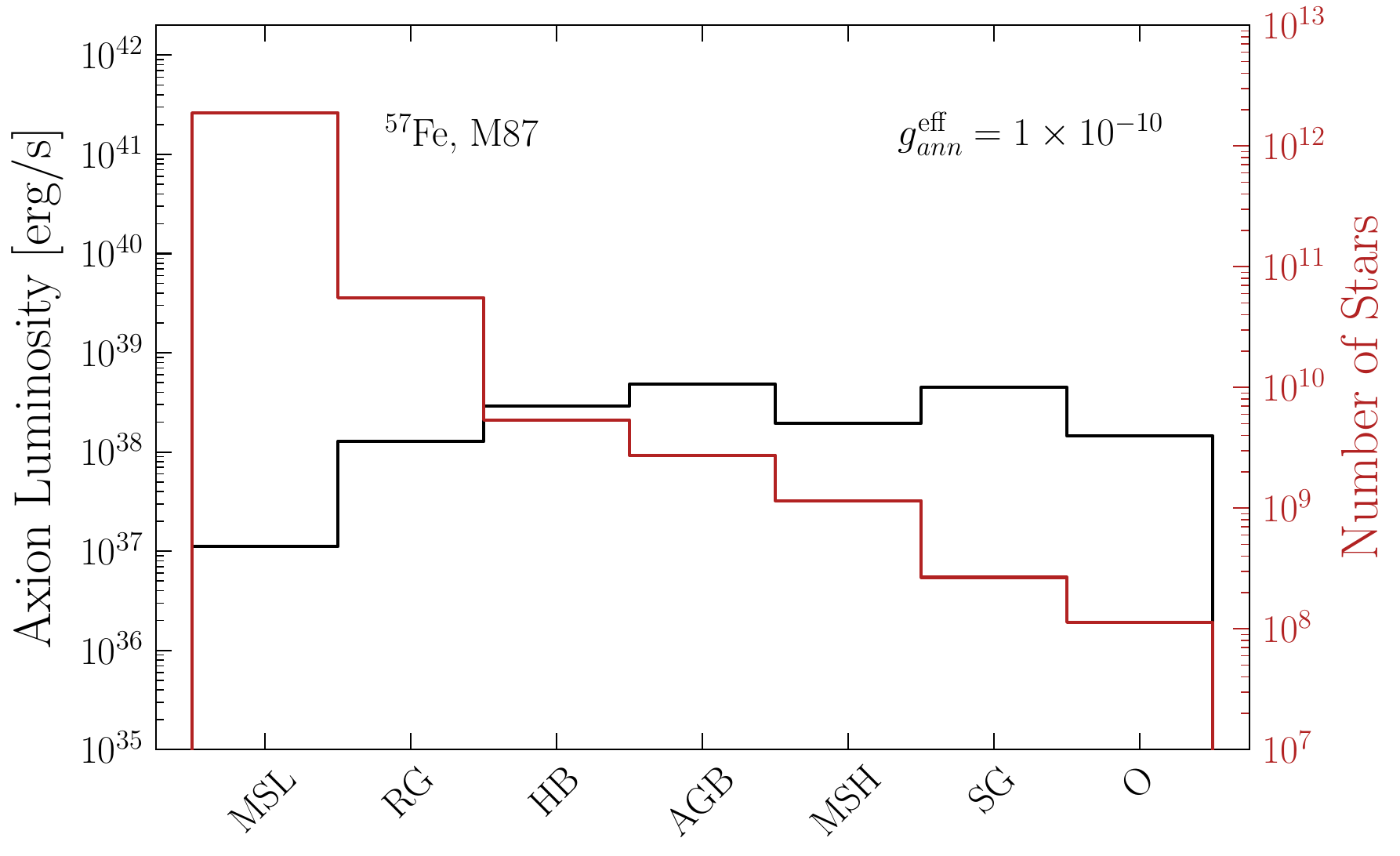}
\vspace{-0.4cm}
\caption{An illustration of the total axion luminosities (black) and stellar numbers (red) across our main stellar subtype populations, for the $\Fe$ transition in M87, at the indicated axion-nucleon coupling. The phases include main sequence for low-mass/high-mass stars (MSL, MSH), red giant (RG), horizontal branch (HB), asymptotic giant branch (AGB), supergiants (SG), and O-type (O) stars.}
\label{fig:stellarsubtype}
\end{figure}

\begin{figure}[!t]
\centering
\includegraphics[width=0.49\textwidth]{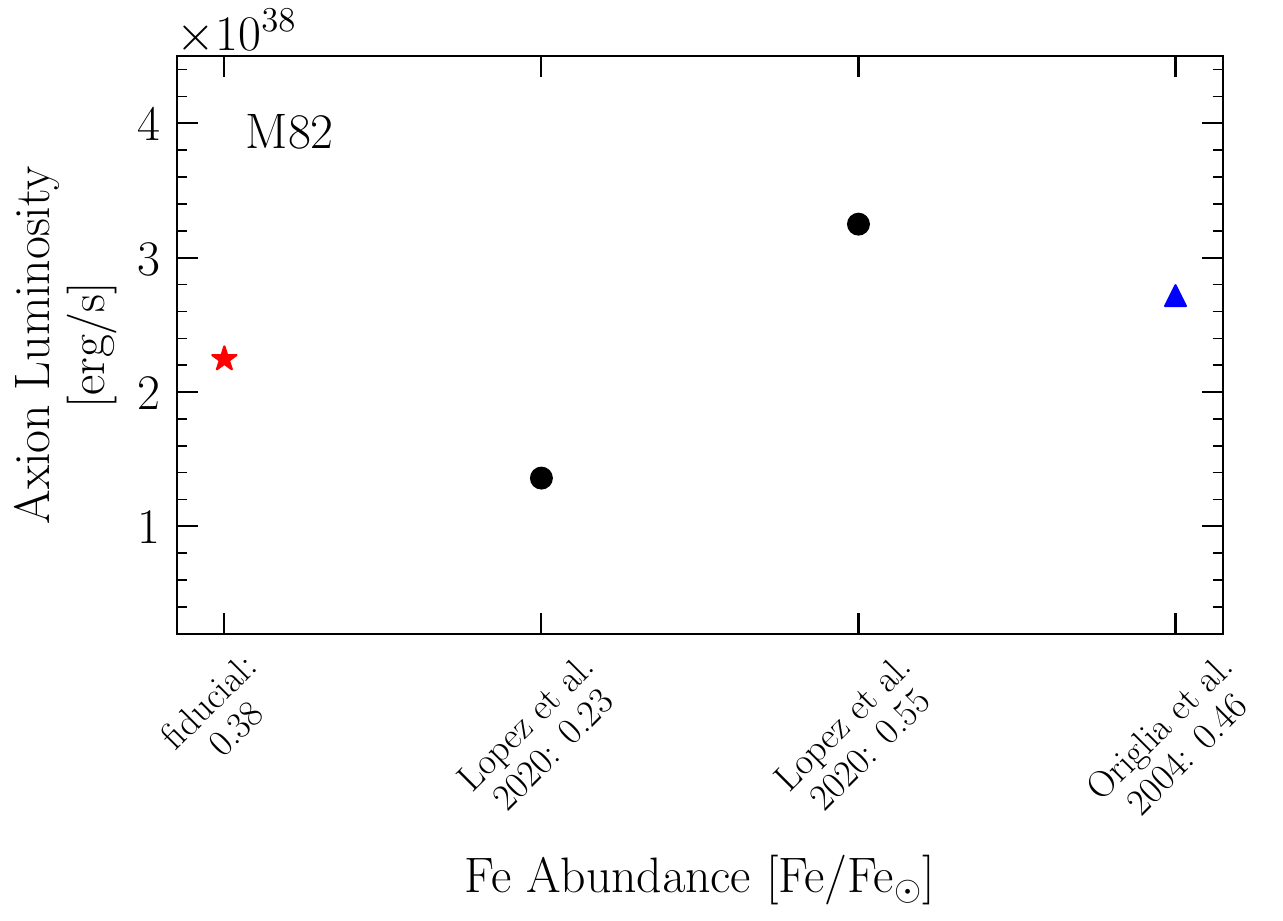}
\vspace{-0.4cm}
\caption{An illustration of the various choices of Fe abundances possible for M82, and their resulting axion luminosities. Shown are the minimum and maximum measured abundances from~\cite{Lopez:2020wxj} (Lopez et al.), as well as the stellar Fe abundance quoted in~\cite{Origlia:2004} (Origlia et al.). The fiducial abundance adopted is taken from the average of the abundances found in~\cite{Lopez:2020wxj}.}
\label{fig:M82_abundances}
\end{figure}

Additionally, we also illustrate in Fig.~\ref{fig:stars_T} the distributions of the core temperatures of the stars comprising the stellar populations of M82 and M87. This gives one a general idea of the accessible energies available for the types of nuclear transition line searches discussed in this work. In Fig.~\ref{fig:evolution}, we illustrate the axion luminosity emitted by an example 20 $M_{\odot}$ star throughout various representative stages of its lifetime. In Fig.~\ref{fig:el_burn} we show the total axion luminosity emitted from the M87 stellar population by burning core element, along with stellar numbers. In Fig.~\ref{fig:stellarsubtype}, also for M87, we show the total axion luminosities and stellar numbers across our main stellar subtype populations. In Fig.~\ref{fig:M82_abundances} we illustrate the effect on axion luminosity from various choices for Fe abundances in M82 (which is likely to be sub-solar, see main text).
Finally, Tab.~\ref{tab:GC_obs} lists the NuSTAR observations used in our analysis of the GC, which are chosen to correspond to the most central region of the bulge, while also keeping away from known bright X-ray transients.

\begin{figure}[!t]
\centering
\includegraphics[width=0.49\textwidth]{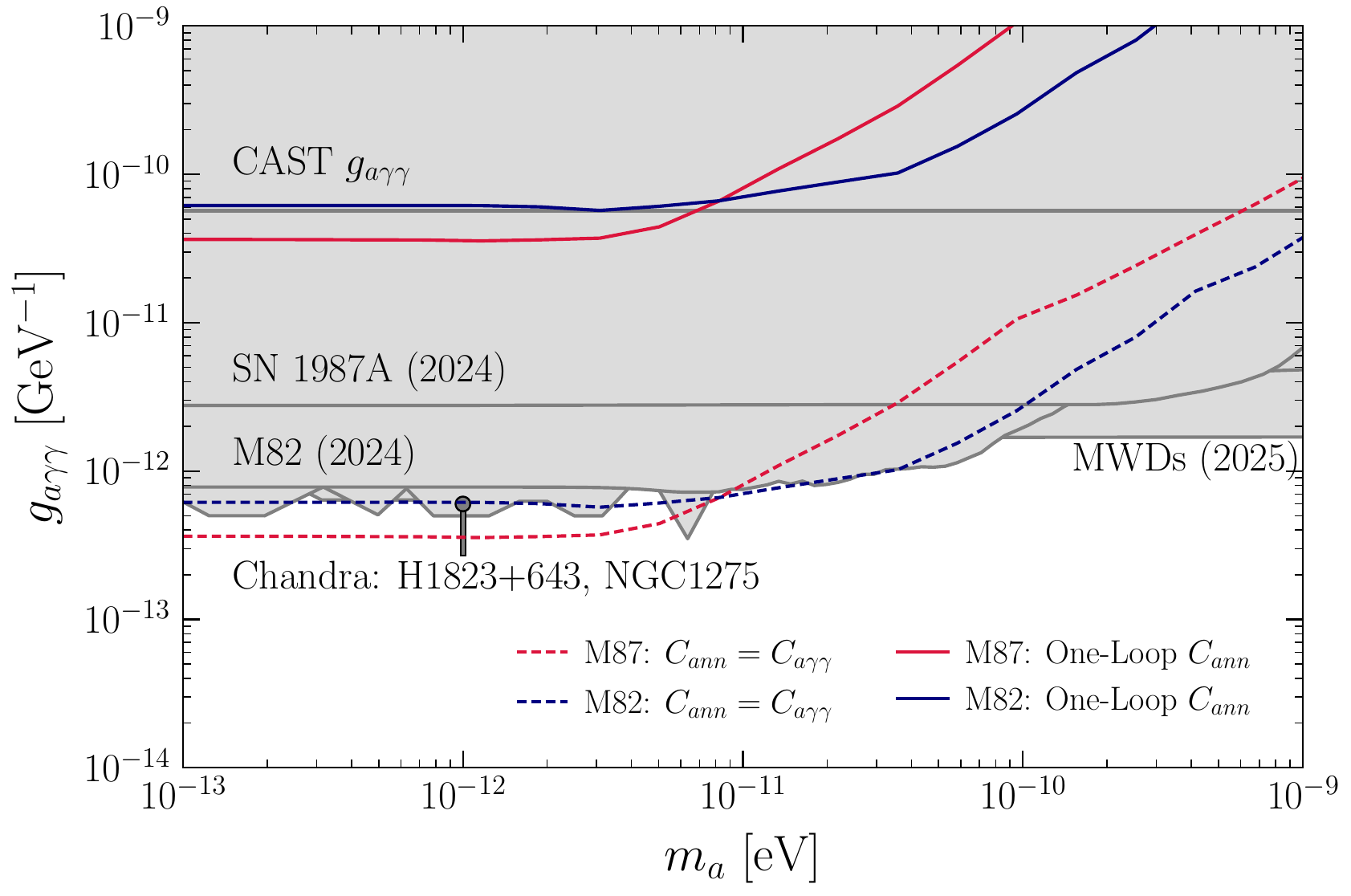}
\vspace{-0.4cm}
\caption{The 95\% limits from Fig.~\ref{fig:ganngagg} translated to limits on $\gagg$ alone under the assumptions of a tree-level axion-nucleon coupling where $C_{ann} = C_{a\gamma \gamma}$, as well as a one-loop induced axion-nucleon coupling where $C_{ann} \sim 10^{-4} C_{a \gamma \gamma}$ (see App.~\ref{app:axionnucleon} for details). Existing limits taken from~\cite{ciaran_o_hare_2020_3932430}.}
\label{fig:gagg_tree_loop}
\end{figure}

\begin{table}[!htb]
\begin{tabular}[t]{>{\centering\arraybackslash}p{2.5cm}>{\centering\arraybackslash}p{2.5cm}}
ObsID & $t_{\rm exp}$ [ks]  \\ \hline \hline
30001002001       & $154.2$       
\\
30001002003       & $77.1$           
\\
30001002004       & $49.6$     
\\

\hline
Total & 280.9
\\
\hline 
\end{tabular}
\caption{\label{tab:GC_obs} Observation IDs and exposure time (in [ks]) of the archival NuSTAR data used in our GC analysis, in which the target is chosen to be the center-most region of the galactic bulge, and the observations are away from known bright X-ray transients.}
\end{table}

\section{Nuclear Transition Calculations and Isotopes}
\label{app:nuclear}

In this section we summarize a survey and comparison of a variety of isotopes besides the ones highlighted in the main text. These additional isotopes with stable nuclear ground states, while not further explored in this work, could be potentially interesting for other related ideas exploring axion production from nuclear de-excitation (\textit{e.g.} such as any search which might only be sensitive to one of $\gann$ or $\gapp$ separately). We leave any such ideas for future work.

First, in Table~\ref{tab:isotopes} we list, in order of increasing excitation energy $E^*$, various selected isotopes whose de-excitation energies could in principle be probed by X-ray axion searches within the range of a hard X-ray telescope such as NuSTAR (\textit{i.e.} with $E^* \lesssim 85$ keV). {We list parameters relevant to axion emission (see~\eqref{eq:nuclear_one},~\eqref{eq:nuclear_two}), such as the angular momenta of the ground and excited states ($J_0$ and $J_1$, respectively), the total mean lifetime $\tau_{m}$, the internal conversion coefficient $\alpha$, the $\delta$ mixing ratio, as well as solar abundances of each isotope and their unpaired nucleons. We note that the precise determination of the nuclear factors $\beta$ and $\eta$ for each nuclei in general should come from detailed calculations of nuclear shell models, which is beyond the scope of this work. However, one can adopt rough approximations of these factors given in~\cite{Massarczyk:2021dje}, which estimates that a nuclei with an unpaired neutron should have values close to $\beta  = -1$ and $\eta = 0.5$, whereas a nuclei with an unpaired proton would instead have $\beta  = 1$ and $\eta = 0.5$. For this reason we include the unpaired nucleon for each isotope as the final column of Table~\ref{tab:isotopes}. We note that while these are only rough estimates, they still serve as useful guidelines for computing branching ratios, \textit{i.e.}~\eqref{eq:fe57_ganneff}, in the absence of precise nuclear shell model calculations, which would be of great interest for future work.}

\begin{figure}[!t]
\centering
\includegraphics[width=0.49\textwidth]{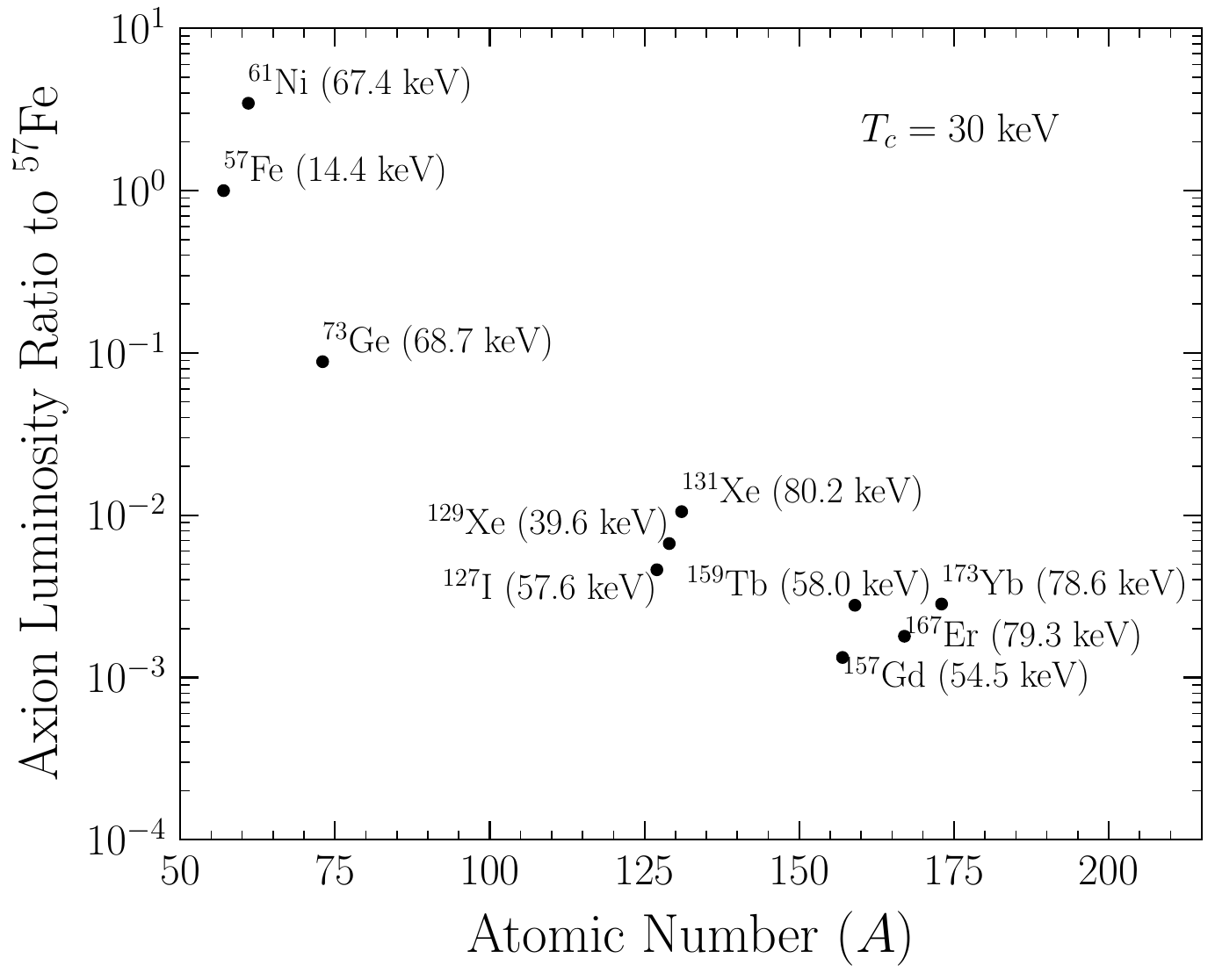}
\vspace{-0.4cm}
\caption{Axion emission rates for various isotopes in a stellar core with a temperature of $T_c = 30$ keV, normalized to the emission rate from $\Fe$. We note that the excitation/de-excitation energies $E^*$ are in parentheses next to the isotope name. For visual clarity, we only select the top ten isotopes with respect to axion luminosity relative to $\Fe$. Here, we see that the three most prominent isotopes are $\Fe$, $\Ni$, and $\Ge$, and so we focus on these isotopes in the main text.}
\label{fig:isotopes_30keV}
\end{figure}

\begin{figure}[!t]
\centering
\includegraphics[width=0.49\textwidth]{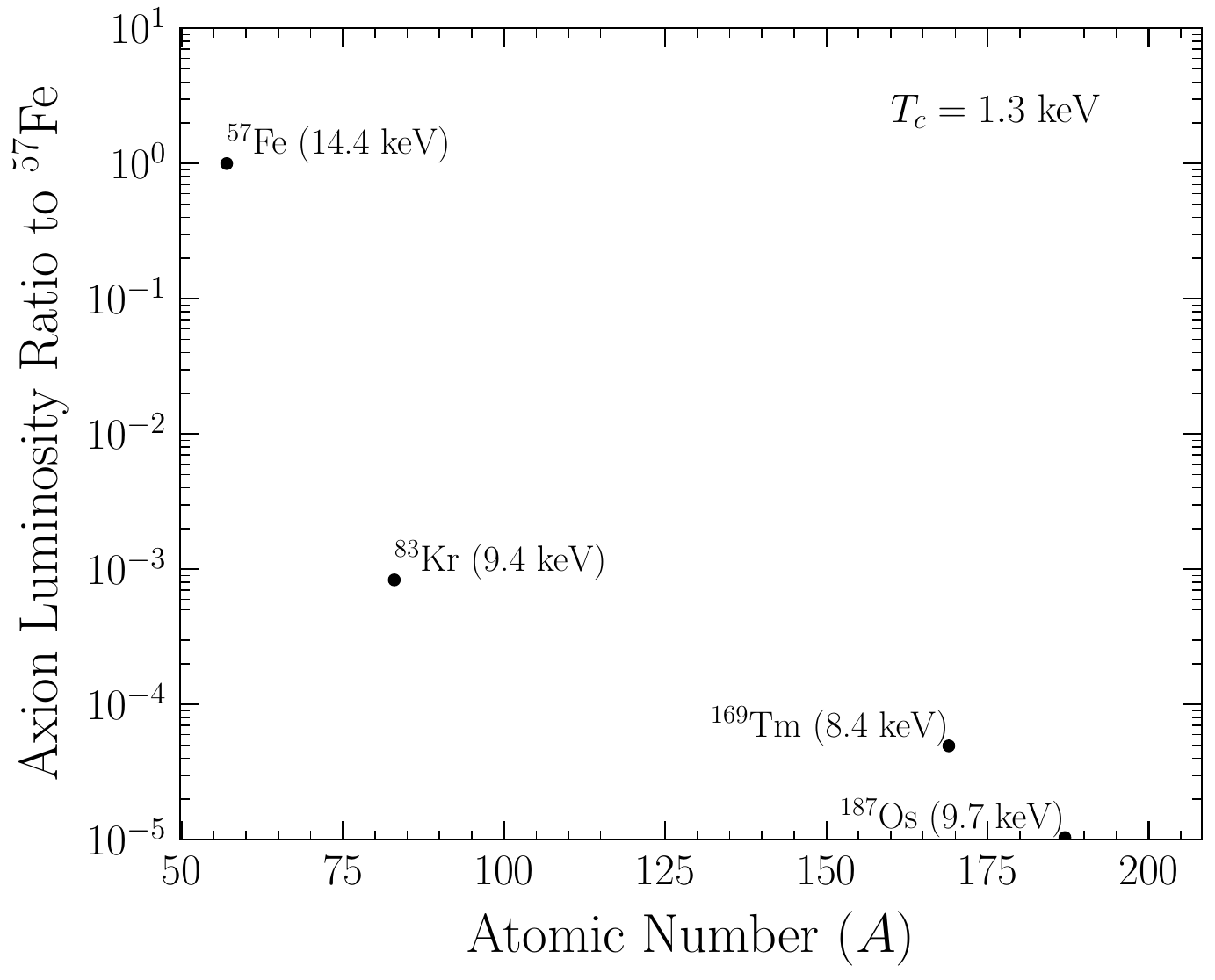}
\vspace{-0.4cm}
\caption{The same as Fig.~\ref{fig:isotopes_30keV} but for a stellar core temperature of $T_c = 1.3$ keV (approximately the core-temperature of the Sun). Only isotopes with contributions above $10^{-5}$ relative to $\Fe$ are shown.}
\label{fig:isotopes_1keV}
\end{figure}

The isotopes in Table~\ref{tab:isotopes} were estimated to be the strongest in terms of axion emission, relevant for either high-mass or low-mass stars, and we compare these isotopes in both stellar cases in Figs.~\ref{fig:isotopes_30keV} and~\ref{fig:isotopes_1keV}, respectively. For the case of a solar-like low-mass star, \textit{i.e.} Fig.~\ref{fig:isotopes_1keV}, it is clear that $\Fe$ is indeed the strongest isotope for axion de-excitation production by a large margin, whereas in the case of a typical high-mass star, \textit{i.e.} Fig.~\ref{fig:isotopes_30keV}, $\Ni$ is a factor of a few stronger, and $\Ge$ is only a little over an order of magnitude weaker than $\Fe$. Hence, we only focus on these isotopes, with particular emphasis on $\Fe$, in our main text. While the numerical values of the de-excitation parameters certainly influence the total axion flux considerably, one general reason why the other numerous isotopes in Table~\ref{tab:isotopes} are largely subdominant in axion flux is because their abundances are far too low in normal stellar matter, which is increasingly true with heavier isotopes. For all of our isotopes, as discussed in the main text, our isotope abundances are primarily informed from galactic metallicities, which we mostly take as roughly solar and hence extract from tables in~\cite{2019arXiv191200844L}.

\begin{table*}[t]
\centering
\begin{tabular}{p{1.8 cm} p{1.8 cm} p{1.6 cm} p{1.6 cm} p{1.6 cm} p{1.6 cm} p{1.6 cm} p{1.6 cm} p{1.6 cm}}
\hline \hline
Isotope & $E_*$ (keV) & $J_0$ & $J_1$ & $\tau_{m}$ (ns) & $\alpha$ & $\delta$ & Solar abundance ($M^{-1}_{\odot}$) & Unpaired nucleon (p/n) \\
\hline
\hline
\\
$^{169}\mathrm{Tm}$ & 8.4 & 1/2 & 3/2 & 5.9   & 263 & 0.033 & $1.4 \times 10^{45}$  & \hspace{0.25 cm} p 
\\
\hline
\\
{$^{83}\mathrm{Kr}$} & 9.4 & 9/2 & 7/2 & 226.2& 17.09
 &0.013 & $2.0 \times 10^{47}$ & \hspace{0.25 cm} n 
\\
\hline
\\
$^{187}\mathrm{Os}$ & 9.7 & 1/2 & 3/2 & 3.43  & 280 & 0.04 &  $2.9 \times 10^{44}$ & \hspace{0.25 cm} n 
\\

\hline
\\
$^{119}\mathrm{Sn}$ & 23.9 & 1/2 & 3/2 & 26  & 5.06 & 0.003 & $1.0 \times 10^{46}$  & \hspace{0.25 cm} n 
\\
\hline
\\
$^{201}\mathrm{Hg}$ & 26.3 & 3/2 & 5/2 & 0.91  & 72.9 & 0.012 & $1.6 \times 10^{45}$  & \hspace{0.25 cm} n 
\\
\hline
\\
$^{125}\mathrm{Te}$ & 35.5 & 1/2 & 3/2 & 2.14   & 13.69 & 0.031 & $1.1 \times 10^{46}$  & \hspace{0.25 cm} n 
\\
\hline
\\
$^{121}\mathrm{Sb}$ & 37.1 & 5/2 & 7/2 & 5   & 10.88 & 0.06 & $6.9 \times 10^{45}$ & \hspace{0.25 cm} p 
\\
\hline
\\
$^{129}\mathrm{Xe}$ & 39.6 & 1/2 & 3/2 & 1.4   & 12.03 & -0.027 & $5.1 \times 10^{46}$  & \hspace{0.25 cm} n 
\\
\hline
\\
$^{183}\mathrm{W}$ & 46.5 & 1/2 & 3/2 & 0.27   & 8.4 & -0.084 & $6.9 \times 10^{44}$  & \hspace{0.25 cm} n 
\\
\hline
\\
$^{157}\mathrm{Gd}$ & 54.5 & 3/2 & 5/2 & 0.19   & 12.1 & 0.19 & $1.8 \times 10^{45}$ & \hspace{0.25 cm} n 
\\
\hline
\\
$^{127}\mathrm{I}$ & 57.6 & 5/2 & 7/2 & 2.81   & 3.72 & -0.083 & $5.4 \times 10^{46}$  & \hspace{0.25 cm} p 
\\
\hline
\\
$^{159}\mathrm{Tb}$ & 58 & 3/2 & 5/2 & 0.08   & 10.73 & 0.119 & $2.1 \times 10^{45}$  & \hspace{0.25 cm} p 
\\
\hline
\\
$^{155}\mathrm{Gd}$ & 60 & 3/2 & 5/2 & 0.28   & 9.14 & -0.198 & $1.7 \times 10^{45}$   & \hspace{0.25 cm} n 
\\
\hline
\\
$^{171}\mathrm{Yb}$ & 66.7 & 1/2 & 3/2 & 1.14   & 12.6 & 0.684 & $1.2 \times 10^{45}$  & \hspace{0.25 cm} n 
\\
\hline
\\
$^{61}\mathrm{Ni}$ & 67.4 & 3/2 & 5/2 & 7.7   & 0.137 & 0.008 & $1.9 \times 10^{49}$  & \hspace{0.25 cm} n 
\\
\hline
\\
$^{73}\mathrm{Ge}$ & 68.7 & 9/2 & 7/2 & 2.57   & 0.238 & 0.074 & $3.1 \times 10^{47}$ & \hspace{0.25 cm} n 
\\
\hline
\\
$^{193}\mathrm{Ir}$ & 73 & 3/2 & 1/2 & 8.79   & 6.1 & -0.558 & $3.5 \times 10^{46}$  & \hspace{0.25 cm} p 
\\
\hline
\\
$^{197}\mathrm{Au}$ & 77.3 & 3/2 & 1/2 & 2.76   & 4.36 & -0.368 & $6.6 \times 10^{45}$  & \hspace{0.25 cm} p 
\\
\hline
\\
$^{173}\mathrm{Yb}$ & 78.6 & 5/2 & 7/2 & 0.07   & 7.01 & -0.224 & $1.4 \times 10^{45}$ & \hspace{0.25 cm} n 
\\
\hline
\\
$^{167}\mathrm{Er}$ & 79.3 & 7/2 & 9/2 & 0.17   & 5.7 & -0.32 & $2.0 \times 10^{45}$ & \hspace{0.25 cm} n 
\\
\hline
\\
$^{131}\mathrm{Xe}$ & 80.2 & 3/2 & 1/2 & 0.69 & 1.57 &  0.1 & $4.0 \times 10^{46}$ & \hspace{0.25 cm} n 
\\
\hline
\\
$^{133}\mathrm{Cs}$ & 81 & 7/2 & 5/2 & 9.06  & 1.7 & 0.158 & $1.2 \times 10^{46}$ & \hspace{0.25 cm} p 
\\
\hline
\\
$^{191}\mathrm{Ir}$ & 82.4 & 3/2 & 1/2 & 5.92  & 10.54 & -0.871 &  $8.0 \times 10^{45}$ & \hspace{0.25 cm} p 
\\
\hline
\\
$^{153}\mathrm{Eu}$ & 83.4 & 5/2 & 7/2 & 1.14  & 3.76 & 0.81 & $5.1 \times 10^{45}$ & \hspace{0.25 cm} p 
\\
    \hline \hline
    \end{tabular}
    \caption{A survey of isotopes potentially interesting for nuclear de-excitation axion line searches with NuSTAR $(E_* \lesssim$ 85 keV). {Nuclear parameter inputs are taken from~\cite{nudat3}, solar abundances are taken from~\cite{2019arXiv191200844L}, and further details, including parameter definitions, are in the main text.}}
    \label{tab:isotopes}
\end{table*}

We finally note that there exist potentially interesting isotopes beyond the X-ray range considered here; for example,  $^{23}\rm{Na}$ was explored in~\cite{Haxton:2025xqz}, and among other isotopes we separately explored, $^{55}\rm{Mn}$, $^{47}\rm{Ti}$, and $^{21}\rm{Ne}$ (with transition energies of order $\mathcal{O}(0.1) - \mathcal{O}(1)$ MeV) seem to potentially have combinations of nuclear parameters and abundances making them marginally comparable, in terms of axion flux, to the isotopes considered in our main work, although for only very narrow choices of stellar masses and lifetimes. The difficulty, furthermore, of transitions on the order of $\mathcal{O}(0.1) - \mathcal{O}(1)$ MeV lies in the reduced sensitivity and general dearth of current telescopes probing these energy ranges.  This was the original reason why an isotope such as $\Fe$ was special; it is an isotope that balances an unusually low transition energy in the $\mathcal{O}(10)$ keV range with a high enough stellar abundance. Upcoming MeV telescopes such as the COSI instrument~\cite{Tomsick:2023aue} would, however, be able to efficiently search for higher energy transitions from other isotopes which could be potentially interesting for axion searches similar to those described in this work. We leave such studies for future work.

\section{Loop-Induced Axion-Nucleon Coupling}
\label{app:axionnucleon}

In this section, we additionally consider a benchmark scenario where our axion-like particles possess couplings generated under the renormalization (RG) flow~\cite{Srednicki:1985xd, Bauer:2017ris, Bauer:2020jbp}. We consider the scenario where the axion only couples to electroweak gauge bosons in the UV, in which case the couplings of axions to the nucleons relevant in this work are induced through RG running. Following the derivation in~\cite{Manzari:2024jns}, and calculating the running at one-loop with a benchmark UV scale of $\Lambda = 10^9$ GeV, one finds that the axion-nucleon coupling coefficients $C_{app}$ and $C_{ann}$ can be derived as
\begin{equation}
\begin{aligned}
    C_{app} &\sim 3.5\times10^{-5} C_{aBB} + 2.1 \times 10^{-4} C_{aWW}, \\
    C_{ann} &\sim -5.6\times 10^{-6} C_{aBB} + 2.1 \times 10^{-4} C_{aWW}.
\end{aligned}
\end{equation}
which means that, for a generic UV completion where we impose $C_{aWW} \sim C_{aBB}$, we find the coupling relations $C_{app}/C_{a\gamma \gamma} \sim C_{ann}/C_{a\gamma \gamma} \sim 10^{-4}$, which we adopt when recasting our original limits on $\ganngagg$ to that of $\gagg$ alone, as in Fig.~\ref{fig:gagg_tree_loop}.

\clearpage
\newpage

\bibliography{refs}

\end{document}